\documentclass[useAMS,usenatbib,graphicx]{mn2e}

\usepackage{graphicx}

\title[HD simulations of feedback in a molecular cloud]{Hydrodynamic simulations of 
 mechanical stellar feedback in a molecular cloud formed by thermal instability}
\author[C. J. Wareing et al.]{C. J. Wareing$^{1}$\thanks{E-mail:
C.J.Wareing@leeds.ac.uk}, J. M. Pittard$^{1}$ and S. A. E. G. Falle$^{2}$\\
$^{1}$School of Physics and Astronomy, University of Leeds, Leeds, LS2 9JT, U.K.\\
$^{2}$School of Mathematics, University of Leeds, Leeds, LS2 9JT, U.K.}
\begin{document}

\date{Accepted 2017 June 6. Received 2017 June 2; in original form 2017 April 8}

\pagerange{\pageref{firstpage}--\pageref{lastpage}} \pubyear{2002}

\maketitle

\label{firstpage}

\begin{abstract}
  We have used the AMR hydrodynamic code, MG, to perform 3D
  hydrodynamic simulations with self-gravity of stellar feedback in a
  spherical clumpy molecular cloud formed through the action of 
  thermal instability. We simulate the interaction of the mechanical
  energy input from 15\,M$_\odot$, 40\,M$_\odot$, 60\,M$_\odot$ and
  120\,M$_\odot$ stars into a 100\,pc-diameter 16,500\,M$_\odot$ cloud
  with a roughly spherical morphology with randomly distributed high
  density condensations.  The stellar winds are introduced using
  appropriate non-rotating Geneva stellar evolution models. In the
  15\,M$_\odot$ star case, the wind has very little effect,
  spreading around a few neighbouring clumps before
  becoming overwhelmed by the cloud collapse.
  In contrast, in the 40\,M$_\odot$, 60\,M$_\odot$ and
  120\,M$_\odot$ star cases, the more powerful stellar winds create
  large cavities and carve channels through the cloud, breaking
  out into the surrounding tenuous medium during the wind phase and
  considerably altering the cloud structure.  After 4.97\,Myrs,
  3.97\,Myrs and 3.01\,Myrs respectively, the massive stars explode as
  supernovae (SNe). The wind-sculpted surroundings considerably affect
  the evolution of these SN events as they both escape the cloud along
  wind-carved channels and sweep up remaining clumps of cloud/wind
  material. The `cloud' as a coherent structure does not survive the
  SN from any of these stars, but only in the 120\,M$_\odot$ case is
  the cold molecular material completely destabilised and returned to
  the unstable thermal phase.
  In the 40\,M$_\odot$ and 60\,M$_\odot$ cases, coherent clumps of
  cold material are ejected from the cloud by the SN, potentially capable of further
  star formation.
\end{abstract}


\section{Introduction}
The radiation fields, winds and SNe of massive stars destroy and
disperse molecular material. This eventually ends star formation in
clusters, though before that happens, massive stars may trigger
further star formation \citep[e.g.][]{koenig12}. The removal of mass
from a cluster affects the cluster dynamics and plays a key role in
cluster dissolution \citep{portegiesZwart10}. Stellar feedback also
sustains turbulence in the ISM \citep[e.g.][]{dobbs11} and powers
galactic fountains and winds \citep{veilleux05}. The mass, momentum,
energy and ionizing radiation fluxes escaping a cluster into a galaxy,
and beyond, depend on how stellar and supernova energy output
regulates cluster gas. However, the coupling of stellar winds, SNe and
ionizing radiation to clumpy, inhomogeneous molecular clouds
surrounding a massive stellar cluster is not well constrained.

Detections of diffuse X-ray emission from many young (pre-SN) massive
star forming regions support the conjecture that winds play important
roles in clusters. Cold molecular material sometimes confines X-ray
emitting gas, but around other clusters, hot gas appears to shape and
remove cold clouds \citep[e.g.][]{townsley14}. Direct evidence for large
scale outflows from stellar clusters is provided by observations of
stellar bowshocks in and near clusters
\citep[e.g.][]{winston12}. ``Leakage'' of the hot gas is also implied by
the much lower mass of hot cluster gas than expected for the cluster
ages and the mass-loss rates of stars \citep{townsley03}, and by
energy budget considerations \citep{rosen14}. This leakage reduces the
thermal pressure within the hot bubble enough that H{\sc ii} gas
pressure may drive the dynamics \citep{harper09}, at least for
molecular clouds of $\sim 10^{5}$\,M$_\odot$
\citep{walch12,dale12}. Simulations of momentum-driven or isothermal
winds \citep{dale15,offner15} give a lower limit to their impact.
Pre-SN feedback enhances the impact of SNe \citep{fierlinger16},
and, in whole galaxy models, clears dense gas from star forming
regions, reducing the star formation rate
\citep[e.g.][]{agertz13}. However, whole galaxy models remain very
sensitive to assumptions in the feedback scheme.

In our previous work \citep{rogers13,rogers14,wareing17} we have
examined the effect that winds and SNe have on surrounding molecular
material.  In \citet[][hereafter Paper II]{wareing17}, we explored the
effect of mechanical feedback from a single 15\,M$_\odot$ star and a
single 40\,M$_\odot$ star into a sheet-like molecular cloud formed by
the action of the thermal instability under the influence of magnetic
fields, that in projection appears remarkably filamentary. That cloud
was one of three cases studied in \citet[][hereafter Paper
I]{wareing16} which explored the formation of dense, cold, molecular
clouds from quiescent diffuse thermally unstable clouds, under the
influence of the thermal instability.  There we considered the
hydrodynamic case, the case of equal magnetic and thermal pressures
and the case of dominating magnetic pressure (10$\times$ greater than
thermal pressure). Paper II explored feedback in the case of equal
pressures. Here, we now explore the hydrodynamic case of feedback into
a roughly spherical clumpy molecular cloud. We have presented a review
of the relevant literature in Paper II and refer the interested reader
back to that work.

In the next section, we present our numerical method and define the
initial conditions used in our model, as well as the basis for
realistic input of mechanical energy from each star. In Section
\ref{results} we present and discuss the resulting simulations,
organised by results for the wind phase, the early SN phase and the
late SN phase, rather than by each star. In Section \ref{analysis} we
present the global evolution with time of energy, density, phase,
mass-weighted temperature-density and pressure-density distribution,
total mass and mixing behaviour in each of our simulations. In Section
\ref{discussion} we compare our results to previous works and relevant
observations. We summarise and conclude the work in Section
\ref{conclusions}.

\section{Numerical Methods and initial conditions}\label{numerical}

\subsection{Numerical methods}

We present 3D, hydrodynamical (HD) simulations of stellar feedback
with self-gravity using the established astrophysical code MG
\citep{falle91} as recently used in Papers I and II. The code employs
an upwind, conservative shock-capturing scheme and is able to employ
multiple processors through parallelisation with the message passing
interface (MPI) library. MG uses piece-wise linear cell interpolation
to solve the Eulerian equations of hydrodynamics. The Riemann problem
is solved at cell interfaces to obtain the conserved fluxes for the
time update. Integration in time proceeds according to a second-order
accurate Godunov method \citep{godunov59}. A Kurganov Tadmor Riemann
solver is again used in this work \citep{kurg00}. Self-gravity is
computed using a full-approximation multigrid to solve the Poisson
equation. We reduce the Magneto-HD code to HD by setting all the
magnetic field components to zero. For full details of the AMR method,
please see Papers I and II. The simulations presented below employed 8
levels of AMR. The physical size of the computational volume and the
physical resolutions are as detailed below.  Continued care has been
taken with the implementation of realistic heating and cooling in the
same way as used in Papers I and II.

\subsection{Initial conditions}

\begin{figure*}
\centering
\includegraphics[width=150mm]{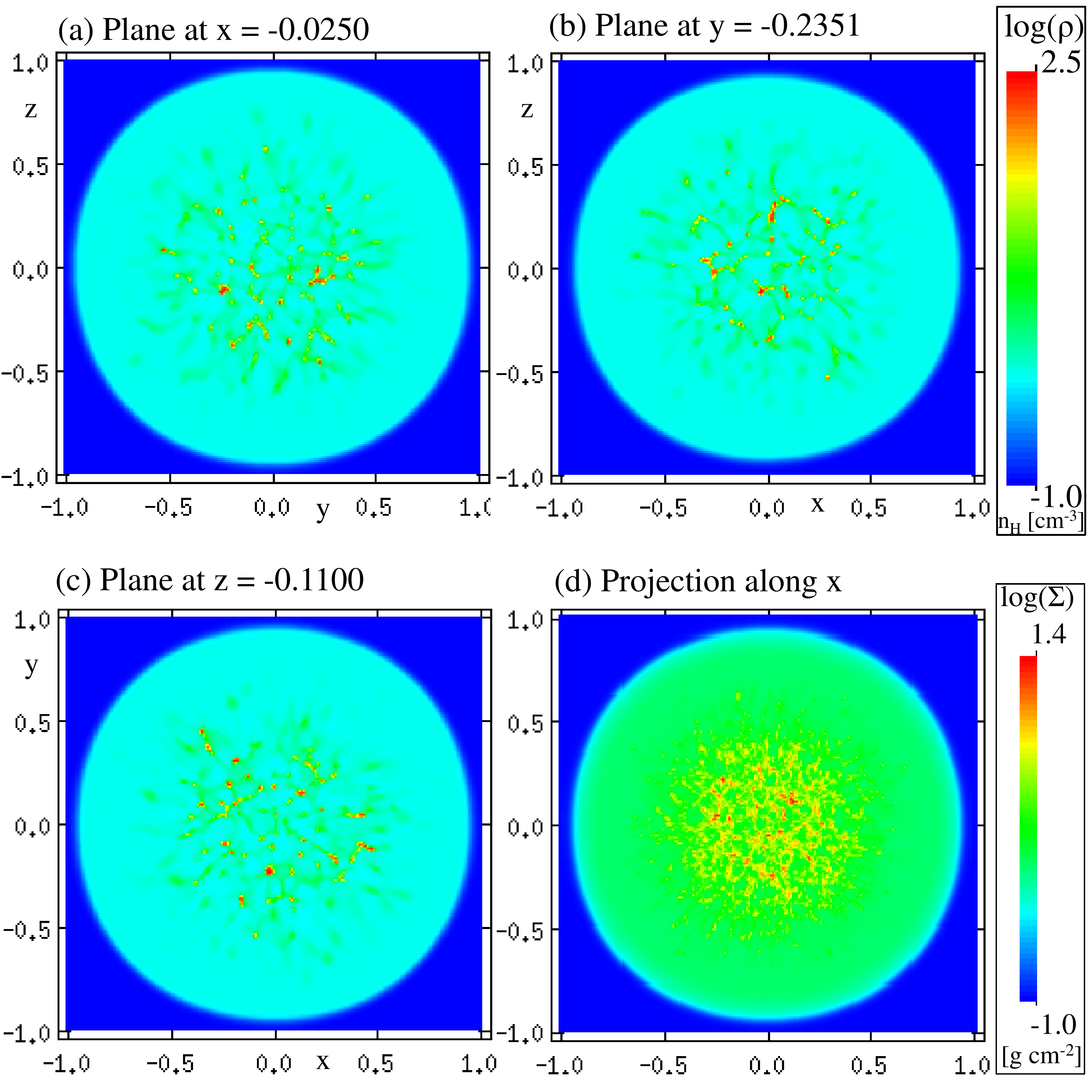}
\caption{Initial condition. Snapshot of a clumpy cloud after 27.1\,Myrs 
of evolution used as the initial condition in this work. Shown are
logarithm of mass density on planes (a) x=-0.025, (b) y=-0.2351, (c) z=-0.11 
and projected column density along the x-axis in (d). 
Length is scaled in units of 50\,pc. Raw data: doi.org/10.5518/201.} 
\label{figure-initial}
\end{figure*}

As well as creating filamentary molecular clouds through the effect of
background magnetic fields and the influence of the thermal
instability \citep{parker53,field65}, Paper I also explored the
formation of clumpy molecular clouds in the zero-field case. In that
work, we examined the evolution of diffuse clouds varying $\beta$, the
ratio of thermal pressure to magnetic pressure.  We examined three
cases, $\beta$ = 0.1, $\beta$ = 1.0 and $\beta$ = $\infty$, equivalent
to the hydrodynamic case of zero magnetic field and of further
interest here. In Paper I, the initial condition consisted of a
stationary cloud of radius 50\,pc with a number density of hydrogen
throughout the cloud of n$_H$ = 1.1 cm$^{-3}$ giving the cloud a total
mass of $\sim$16,500\,M$_\odot$. In the cloud, 10\% density variations
about the uniform initial density were introduced. The pressure was
set according to the local density and thermal equilibrium between
heating and cooling prescriptions at P$_{eq}$/k = $4700\pm300$
K\,cm$^{-3}$, resulting in an initial temperature T$_{eq}$ =
$4300\pm700$\,K (an unstable part of the equilibrium curve - for more
details see Paper I). An advecting scalar, $\alpha_{cloud}$,  was set equal
to 1 in the cloud material. The pressure of the lower density (n$_H$ = 0.1
cm$^{-3}$) surroundings was set equal to that of the unperturbed
cloud, with $\alpha_{cloud}$ set equal to zero. 
No magnetic field was present in the simulation.

The computational volume consists of a 150\,pc$^3$ box with free-flow
boundary conditions (non-periodic for self-gravity) and AMR level G0
containing 4$^3$ cells. Eight levels of AMR mean an effective
resolution of 512$^3$ cells on level G7, although by the time of the
initial condition G2 with 16$^3$ cells is the finest fully populated
level (rather than the MG default of G1). Extra tests have shown we
have resolved the action of the thermal instability. As previously
noted in Papers I and II, such a large number of AMR levels is
employed in order to efficiently compute the self-gravity on the
coarsest levels and also fully resolve the structures formed in the
molecular cloud. The finest physical resolution is 0.293\,pc.  The total 
number of cells across all eight AMR levels is $13.9\times10^6$ grid 
cells, ten times less than that required
by a fixed grid code with the equivalent 512$^3$ cells. The
large number of AMR levels is computationally costly. Each 3D HD
simulation with stellar feedback and self-gravity presented here with
8 AMR levels took approximately 60,000 CPUhours ($\sim$10-12 48-hour
cycles on 128 cores of the high performance computing facility
at Leeds), so 240,000 CPUhours for the four runs presented. Supporting
investigations included a parameter exploration of the initial
condition (see Paper I) and a high resolution re-run of the purely
hydrodynamic case which also explored clump-collapse to form
pre-stellar cores. This cost more than the total CPUhours used by the
simulations presented herein. Confidence in these simulations also
comes from the results presented in Paper II. Each further level of
AMR introduces a computational-cost-multiplier of between 6 and 8 for
this model, given the spherical nature of the cloud, thus making
complete higher resolution simulations prohibitively expensive at this
time. Tests in Paper II showed the resolution used is appropriate for
these simulations, though it is close to the cooling length at times.

The influence of the thermal instability causes the cloud to evolve
into multiple clumps. Specifically,
molecular clumps form throughout the diffuse cloud.
Self-gravity accelerates the contraction of the cloud.
Besides exploring the importance of the thermal
instability in molecular cloud evolution, the secondary aim of Paper I
was to provide a more realistic initial condition for this work, by
including more accurate heating and cooling, the effect of thermal
instability, self-gravity and magnetic fields, as compared to our
previous feedback studies \citep{rogers13,rogers14} where
the clump structure was based upon the simulations of
\cite{vazquez08} of turbulent and clumpy molecular clouds and
contained 3240\,M$_{\odot}$ of material in a 4\,pc radius.

\begin{figure*}
\centering
\includegraphics[width=150mm]{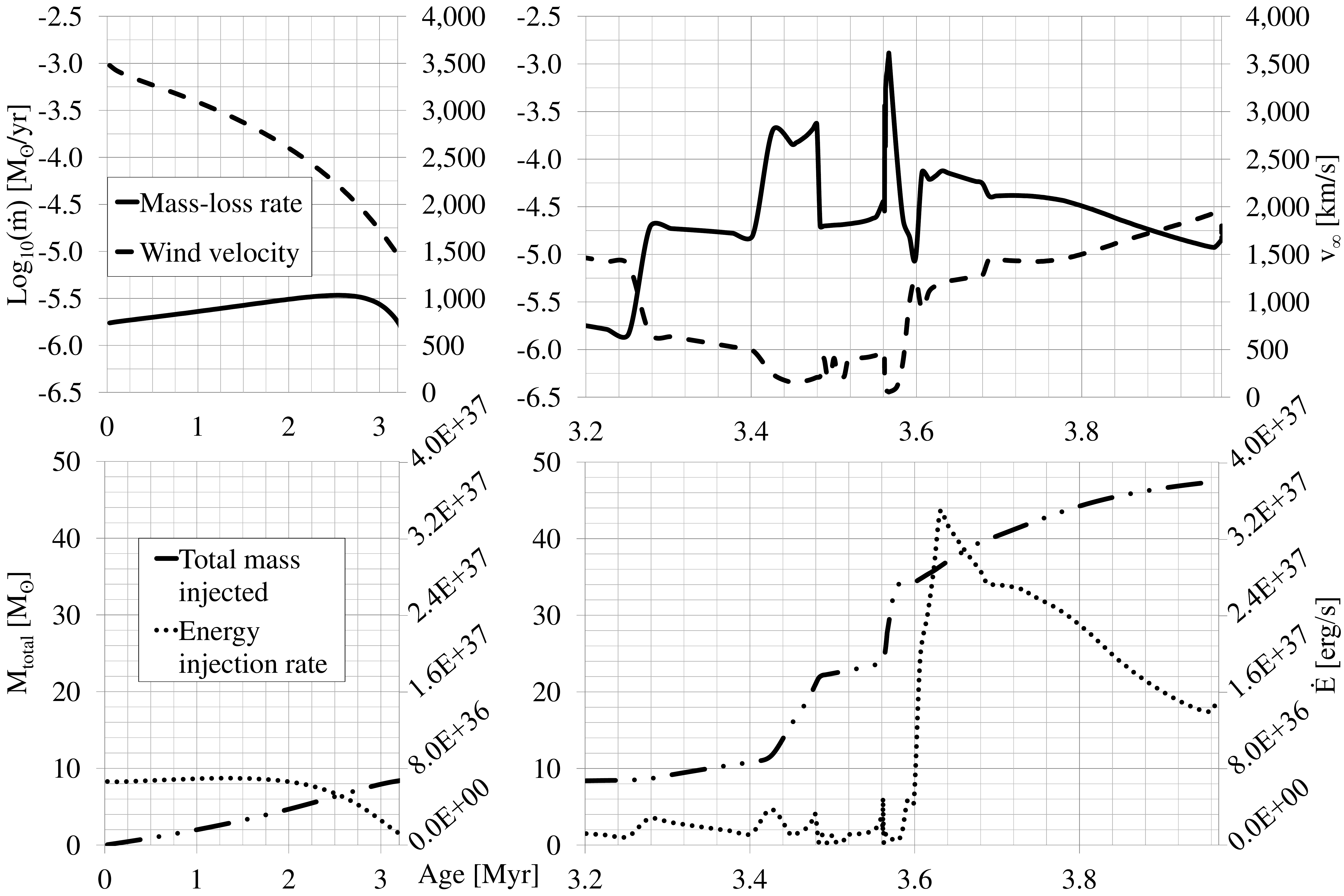}
\caption{Stellar evolution tracks \citep{vink00,vink01,ekstrom12} for 
a 60\,M$_\odot$ star, showing mass-loss rate and wind velocity on the upper graphs, and energy 
injection rate and total injected mass on the lower graphs. Raw data: doi.org/10.5518/201.} 
\label{60evolution}
\end{figure*}

In this work, we take as our initial condition a repeat of a
cloud simulation without magnetic field following the method of Paper
I. Different random seeds result in a simulation that is qualitatively
the same as the result in Paper I, but quantitatively
different.  After 27.1\,Myrs of evolution, densities in the
condensations formed in this new simulation have reached
100\,cm$^{-3}$ - the density threshold often used for injection of
stars in similar simulation work \citep[e.g.,][]{fogerty16}. In this
work, we have decided to inject stars at this time following the work
of such other authors and also because of our own high-resolution
simulations.  
We show snapshots of this initial condition
in Figure \ref{figure-initial} and refer the interested reader to
Paper I for a full description of the evolutionary process that led to
the formation of this cloud. It is important to note though that this
time-scale of 27.1\,Myrs should not be considered as the `age' of the
parent molecular cloud - this is the length of time required to go
from a diffuse cloud with an average density of n$_H$ = 1.1 cm$^{-3}$
to a structured molecular cloud where feedback can be introduced.
Cold (less than 100\,K) condensations in the cloud have only existed
for a few Myrs, in reasonable agreement with observed ages of
molecular clouds.

In this paper, we consider four scenarios, each employing this initial
condition, in order to examine the effect of stellar feedback in this cloud.
We use the same method as Paper II, but repeat ourselves here for 
clarity in this work.

\begin{figure*}
\centering
\includegraphics[width=150mm]{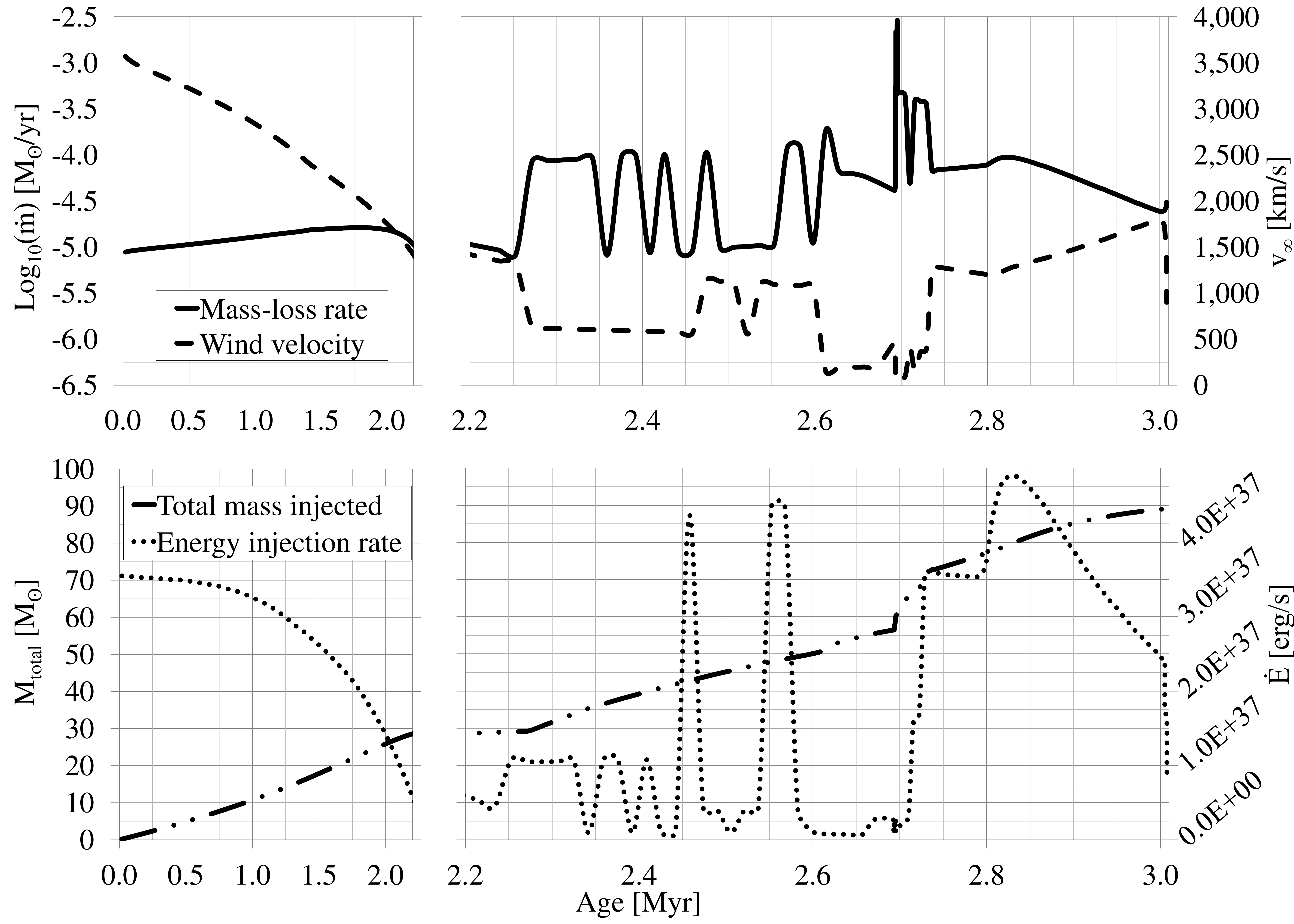}
\caption{Stellar evolution tracks \citep{vink00,vink01,ekstrom12} for 
a 120\,M$_\odot$ star, showing 
mass-loss rate and wind velocity on the upper graphs, and energy 
injection rate and total injected mass on the lower graphs.
Raw data: doi.org/10.5518/201.} 
\label{120evolution}
\end{figure*}

\subsubsection{Scenario 1 - a 15\,M$_{\odot}$ star}

In this scenario we introduce a 15\,M$_\odot$ star at the position
($x$,$y$,$z$) = (-0.025, -0.2351, -0.11) where the coordinates are 
given in scaled code units and the grid extends to $3\times3\times3$ 
centred on (0, 0, 0). This is the location of the highest
density condensation in the cloud, closest to the centre of the
volume. We remove enough mass present in a spherical volume with a 5-cell
radius centred at this point to form a 15\,M$_\odot$ star, assuming
100\% conversion of cloud material to star. For this first
investigation, the cloud mass in this spherical region is removed at the
switch on of the stellar wind, t$_{wind}$=0, under the assumption
that this material has formed the star. If the mass is left in the 
injection region, the stellar
wind rapidly and unrealistically cools and hence the feedback effects
are significantly underestimated. An advected scalar, $\alpha_{wind}$,
previously set to zero throughout the grid, is set to 1 in the wind
injection region in order to track the movement and mixing of the wind
material.

For the stellar evolution, a 15\,M$_{\odot}$ non-rotating Geneva
stellar evolution model calculated by \cite{ekstrom12} is used in
order to provide a realistic mass-loss rate over the lifetime of the
star, as per the method used in Paper II. Detailed plots of this 
evolution can be found in Paper II. In this case, after 12\,Myrs
of stellar evolution, the cloud has also collapsed under the influence
of gravity. Densities in the centre of the cloud have reached levels
where the resolution is insufficient so the simulation was stopped at
this point, close to the end of the star's lifetime of 12.5\,Myrs.
During the wind phase, the total mass and total energy injected by the
star are 1.75\,M$_\odot$ and $1.05\times10^{49}$\,erg respectively.
The location of the star remains constant throughout this simulation.
The centre of cloud-collapse is not exactly at the location of the
star though, so in this case of a 15\,M$_\odot$ star, a moving source
is strictly required to accurately model this case. In future work, we
plan to convert the mass into a `star' particle following the method
in MG of \cite{vanloo13} and \cite{vanloo15}.  The star can then move
through the computational volume whilst feeding back through winds and
SNe and remain consistent with self-gravity in the simulation.

\subsubsection{Scenario 2 - a 40\,M$_{\odot}$ star}

In this scenario we introduce a 40\,M$_\odot$ star using the same
method and at the same position as in Scenario 1, removing enough
mass to form a 40\,M$_{\odot}$ star from the spherical injection
region. A 40\,M$_{\odot}$
non-rotating stellar evolution model calculated by \cite{ekstrom12} is
used in order to provide a realistic mass-loss rate over the lifetime
of the star.  Detailed plots of this evolution can be found in Paper
II. The total mass and total energy injected by the star prior to
supernova explosion are 27.2\,M$_\odot$ and $2.50\times10^{50}$\,erg
respectively. After 4.97\,Myrs, the star explodes as a SN, injecting
10\,M$_\odot$ of stellar material and 10$^{51}$ ergs of energy into
the same wind injection volume. The SN mass and energy is injected
over 500\,yrs, roughly consistent with the time taken for a remnant to
reach the size of the injection volume. An advected scalar,
$\alpha_{SN}$, previously set to zero throughout the grid, is set to 1
in the supernova injection region in order to track the movement and
mixing of the SN material. At this time the wind scalar
$\alpha_{wind}$ is set to zero. The fraction of cloud material in any
given cell is $\alpha_{cloud} - \alpha_{wind} - \alpha_{SN}$.

Gravity plays less of a role during this shorter stellar wind phase 
as compared to the 15\,M$_\odot$ star case and the early SN
phase, partly due to the comparatively powerful dynamics and partly
due to the shorter timescale, but we continue to include it
for consistency within the cloud and to explore the evolution
post-SN when it again plays more of a role. 

\subsubsection{Scenario 3 - a 60\,M$_{\odot}$ star}

In this scenario we introduce a 60\,M$_\odot$ star using the same
method and at the same position as in Scenario 1, removing enough
mass to form a 60\,M$_{\odot}$ star from the spherical injection
region. A 60\,M$_{\odot}$ non-rotating 
stellar evolution model calculated by \cite{ekstrom12} is used in 
order to provide a realistic mass-loss rate over the lifetime of the star.
The calculated mass-loss rate and wind velocity are shown in
Fig.~\ref{60evolution}. Also shown in Fig.~\ref{60evolution} are the energy
injection rate and total injected mass. 
The total mass and total energy injected by the star prior to supernova 
explosion are 47.5\,M$_\odot$ and $8.9\times10^{50}$\,erg respectively,
almost equivalent to the energy introduced in the SN event and considerably
more material. After
3.97\,Myrs, the star explodes as a SN in the same manner as in 
Scenario 2.

\subsubsection{Scenario 4 - a 120\,M$_{\odot}$ star}

In this scenario we introduce a 120\,M$_\odot$ star using the same
method and at the same position as in Scenario 1. In this case, we simply
remove all the mass in a spherical region twice the size of the injection 
region in order to form the 120\,M$_\odot$ star. A 120\,M$_{\odot}$ non-rotating 
stellar evolution model calculated by \cite{ekstrom12} is used in 
order to provide a realistic mass-loss rate over the lifetime of the star.
The calculated mass-loss rate and wind velocity are shown in
Fig.~\ref{120evolution}. Also shown in Fig.~\ref{120evolution} are the energy
injection rate and total injected mass. The total mass and total
energy injected by the star prior to supernova explosion are
89.1\,M$_\odot$ and $2.44\times10^{51}$\,erg respectively,
considerably more than the amount of material and energy 
introduced in a SN event. After
3.01\,Myrs, the star explodes as a SN in the same manner as in 
Scenario 2.

\begin{figure*}
\centering
\includegraphics[width=160mm]{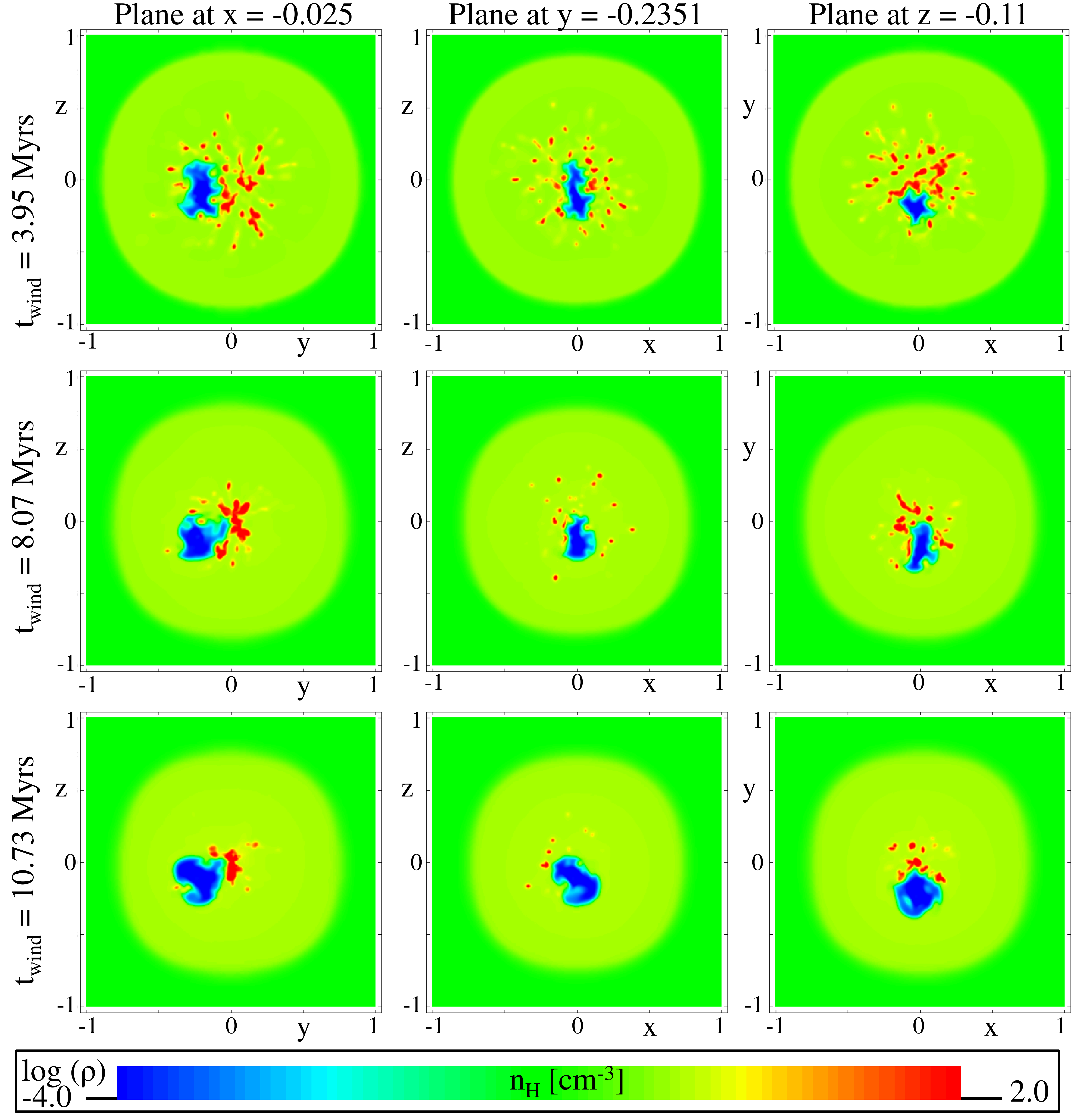}
\caption{Cloud-wind interaction during the lifetime of a 15\,M$_\odot$ star. 
Shown is the logarithm of mass density at various times and on various 
planes through the computational volume. Length is scaled in units of 50\,pc.
Raw data: doi.org/10.5518/201.} 
\label{wnd-15M}
\end{figure*}

\section{Results}\label{results}

In this section we present our results. We present both 2D slices
through the computational volume created within MG and 3D contour and
volume visualisations, created using the VisIt software \citep{visit}.
Raw data for the all figures in this paper is available
from the University of Leeds Repository at doi.org/10.5518/201.

\subsection{The wind phase}\label{windphase}

\begin{figure*}
\centering
\includegraphics[width=160mm]{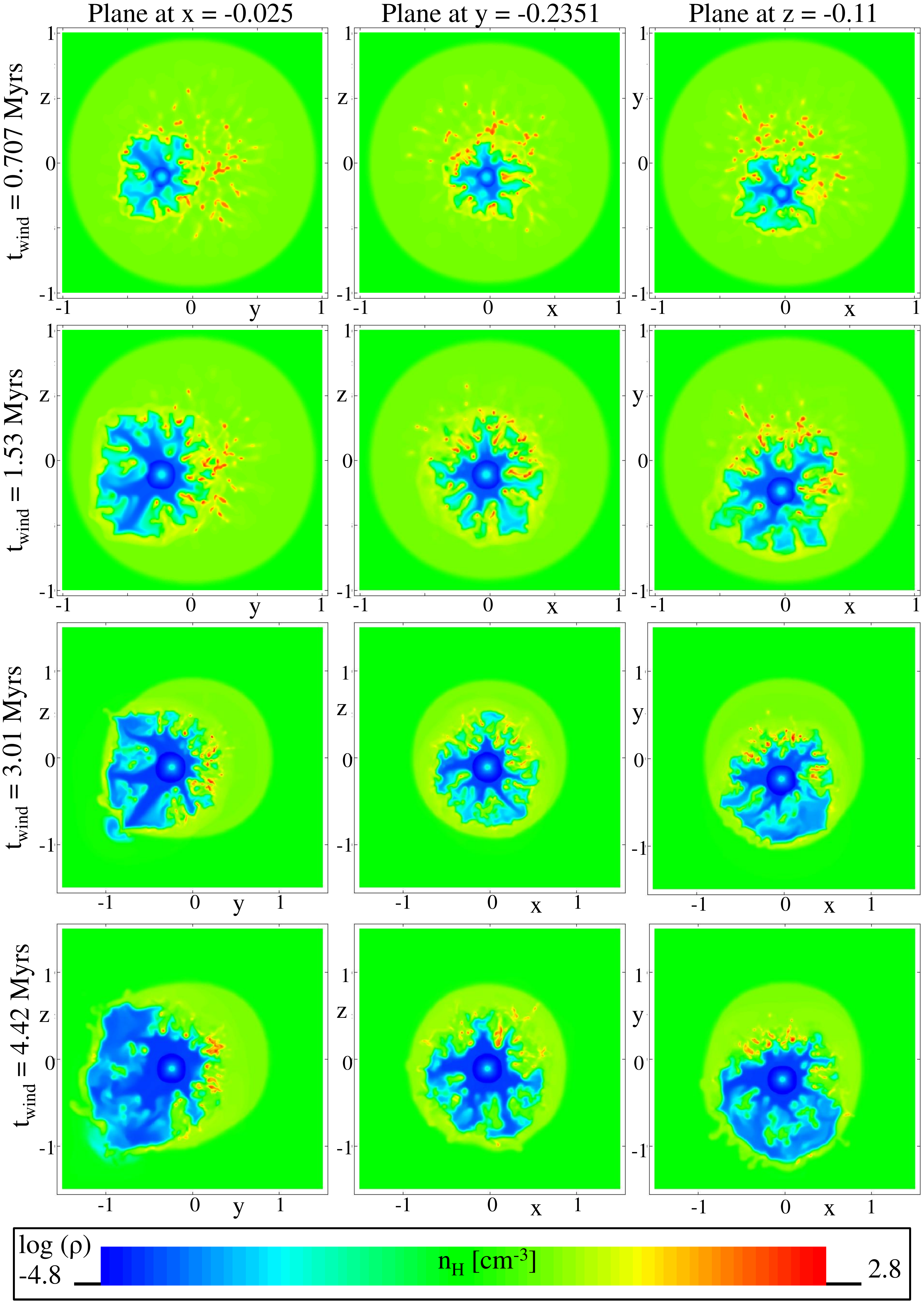}
\caption{Cloud-wind interaction during the lifetime of a 40\,M$_\odot$ star. 
Shown is the logarithm of mass density at various times and on various 
planes through the computational volume. Length is scaled in units of 50\,pc.
Raw data: doi.org/10.5518/201.} 
\label{wnd-40M}
\end{figure*}

\begin{figure*}
\centering
\includegraphics[width=160mm]{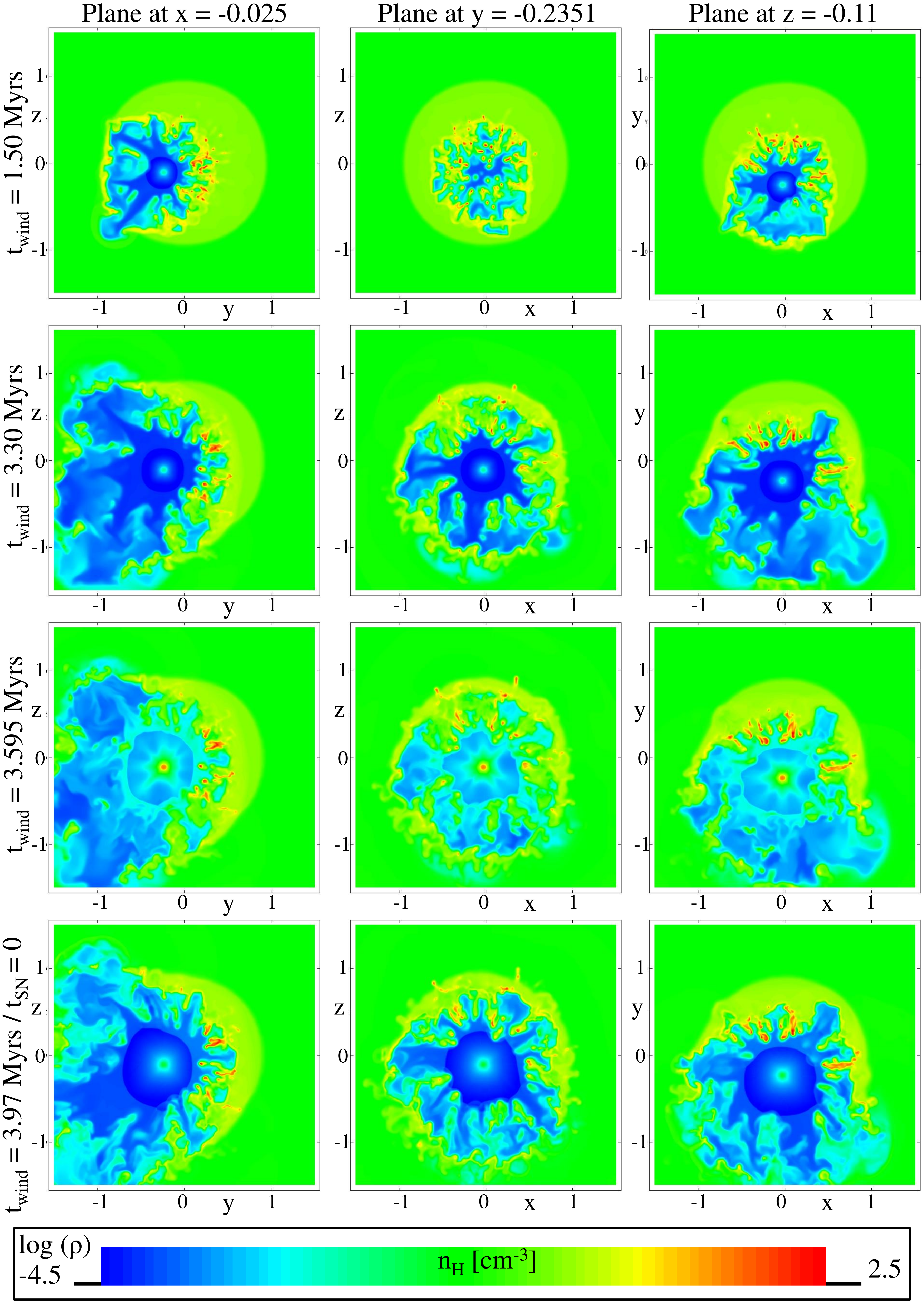}
\caption{Cloud-wind interaction during the lifetime of a 60\,M$_\odot$ star. 
Shown is the logarithm of mass density at various times and on various 
planes through the computational volume. Length is scaled in units of 50\,pc.
Raw data: doi.org/10.5518/201.} 
\label{wnd-60M}
\end{figure*}

\begin{figure*}
\centering
\includegraphics[width=160mm]{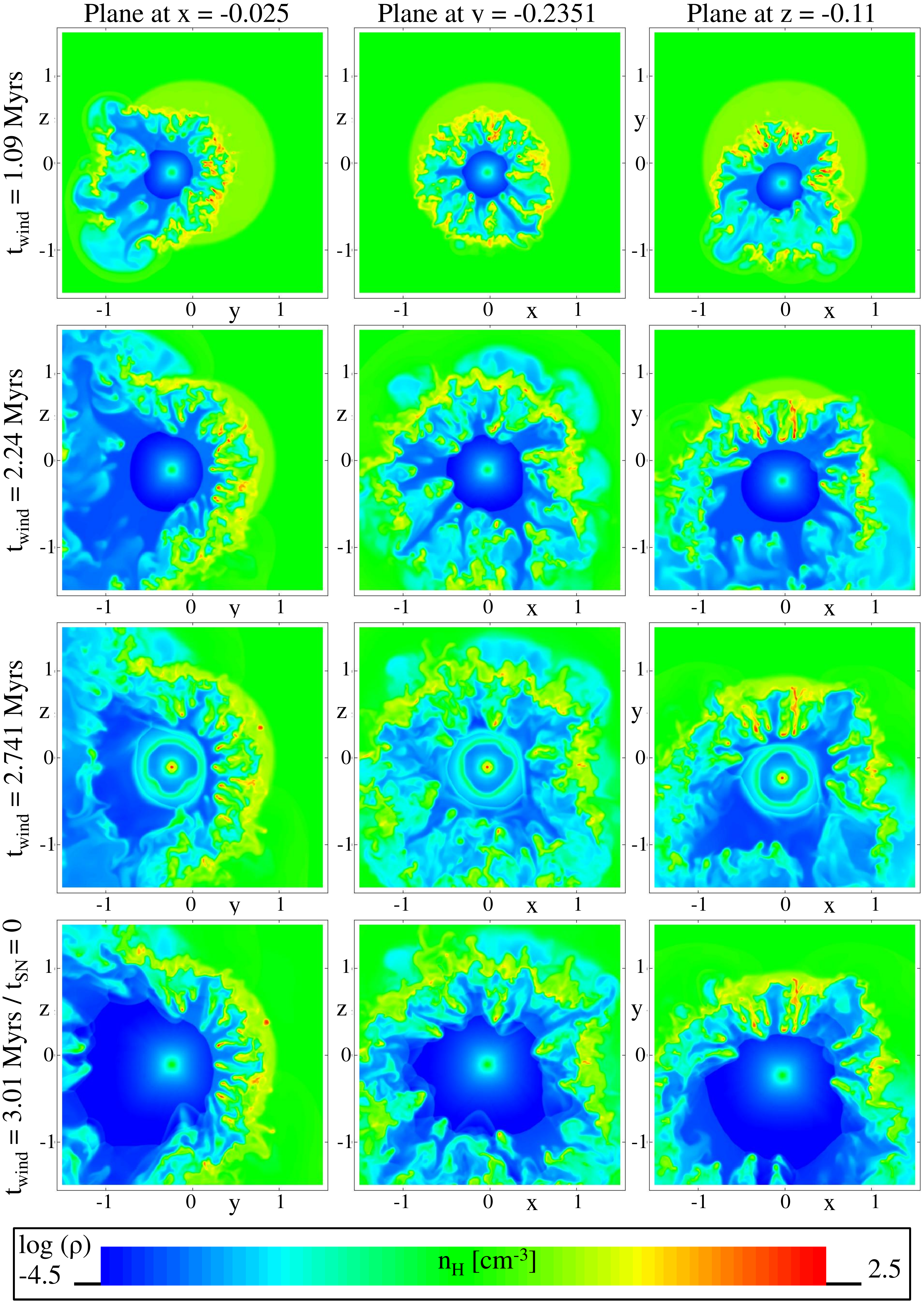}
\caption{Cloud-wind interaction during the lifetime of a 120\,M$_\odot$ star. 
Shown is the logarithm of mass density at various times and on various 
planes through the computational volume. Length is scaled in units of 50\,pc.
Raw data: doi.org/10.5518/201.} 
\label{wnd-120M}
\end{figure*}

\begin{figure*}
\centering
\includegraphics[width=160mm]{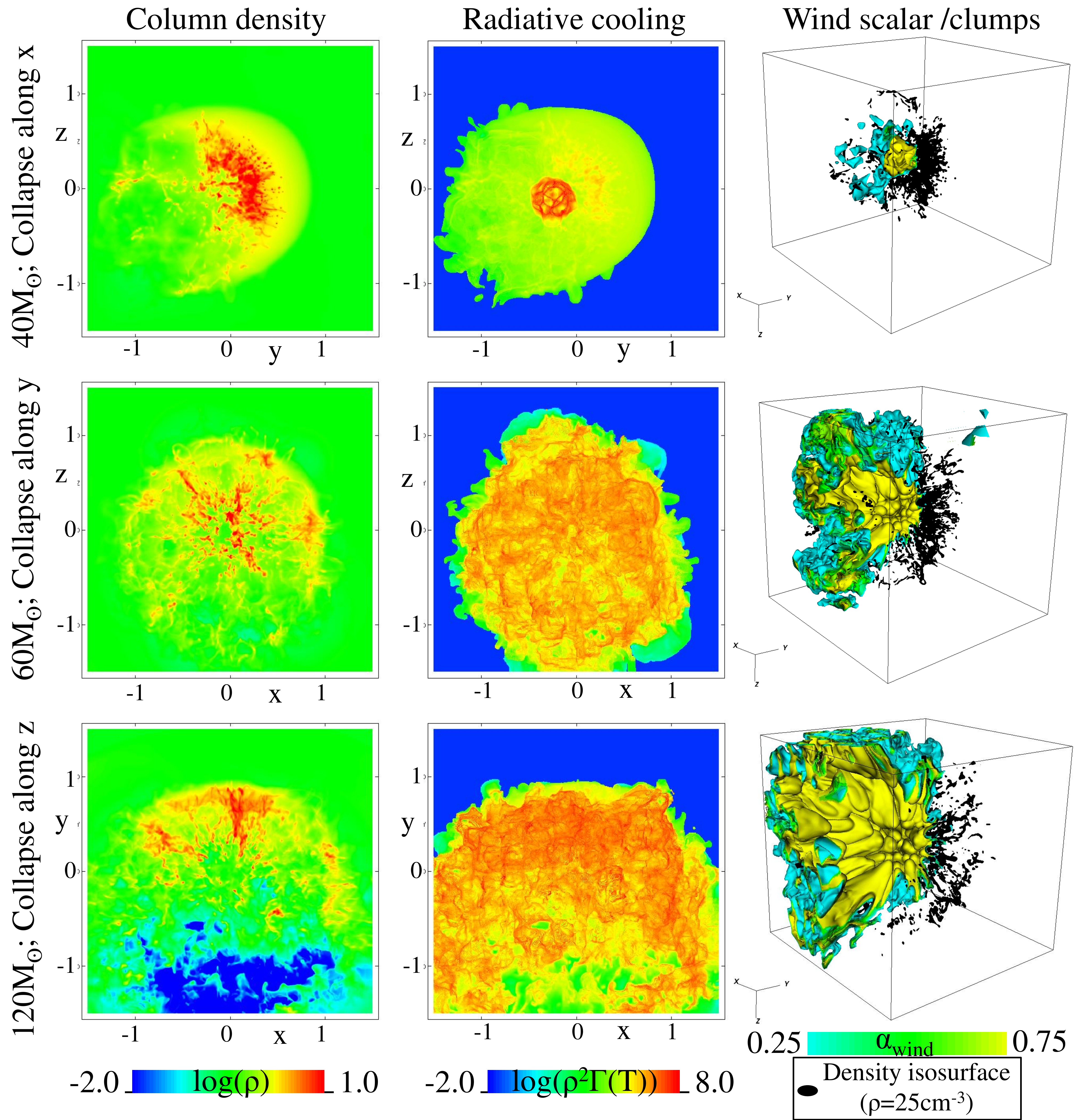}
\caption{Column density, naive emission and 3D visualisation of the cloud-wind
structure just before each star goes SN. Naive emission is calculated as the
radiative cooling energy source for each cell at this time, projected along the
axes as defined in the figure. Length is scaled in units of 50\,pc. Each 3D
visualisation shows the isosurface of the wind scalar (in colour) in half the
computational volume ($x\leq0$) in order to show the core of the wind-blown
bubble and the distribution of high density clumps (black isosurfaces) in the 
entire computational volume. Raw data: doi.org/10.5518/201.} 
\label{wnd-proj}
\end{figure*}

In Fig.~\ref{wnd-15M} we show density slices through the computational 
volume during the evolution of the 15\,M$_\odot$ star up to the end of its 
life. The low mass-loss rate and corresponding low energy injection rate have
minimal effect on the cloud structure, generating only a wind cavity that
spreads away from the star into the cloud around the clumps for the first
few Myrs, but is eventually confined by the gravitational contraction of the
cloud itself. All three planes through the location of the star reveal similar
structure for all timepoints. This structure is a result of the weak wind that
is able to affect the inter-clump material, but is not able to 
affect the high-density clumps themselves. Examining the extent of the 
stellar wind material we see that very little wind material gets more than 
10\,pc away from the star.

During the red super giant (RSG) phase of evolution, from
t$_{wind}$=11.2\,Myr until its SN explosion, the slow dense wind
deposits considerable amounts of material into the cavity formed by
the earlier wind and completely refills it. We do not show this stage
of evolution here, as the cloud itself has collapsed and dominates the
end stages of the stellar bubble's evolution. Neither do we trigger a
supernova in this simulation, as the cloud collapses to a high density
which is unrealistic given our resolution.

In Fig.~\ref{wnd-40M} we show the logarithm of density on all three
planes through the location of the 40\,M$_\odot$ star at various times
through the star's lifespan. By only 0.707\,Myrs into the main
sequence evolution of the star, the impact on the molecular cloud is
significant and clearly different from the 15\,M$_{\odot}$ star case,
as shown in the first row of Fig.~\ref{wnd-40M}. The stellar wind is
expanding away from the star, streaming past the clumps and already
forming clear channels through the cloud. The wind is also beginning
to ablate material from the clumps into the wind flow along the
channels. By 1.53\,Myrs, as shown in the second row of
Fig.~\ref{wnd-40M}, the wind has continued to expand, establishing a
distinct reverse shock at a radius of approximately 5\,pc.  The
ablated clumps are losing further material and being pushed away from
the star - the force of the stellar wind on the clumps is clearly
overcoming the gravitational contraction of the cloud and the wind is
now blowing the cloud apart. After 3.01\,Myrs, the wind has reached
the edge of the cloud material and can be seen in the
$x$-plane expanding into the low-density surrounding medium. The wind
has not yet escaped the cloud on the other planes, probably
due to a combination of the centre-offset position of the star and the
clump distribution in the cloud.  A small number of low-density
channels have now also become dominant in the structure, but by
4.42\,Myrs, at the end of the Main Sequence evolution of the star,
these channels have become less distinct on the x-plane, but more
distinct on the $y$-plane. This structure, with its channels and
clumps connected together is reminiscent of our previous work
\citep{rogers13}.

The star now enters the LBV phase. The wind mass-loss rate increases
by two orders of magnitude to $\sim10^{-4}$\,M$_\odot$\,yr$^{-1}$, and
the terminal wind speed reduces to $\sim100$\,km\,s$^{-1}$. This slow
dense wind forms a high density environment around the location of the
star, which eventually contains $\sim20$\,M$_\odot$ of LBV wind
material.  This phase lasts approximately 200\,kyrs and is followed by
the WR phase of stellar evolution, where a variable, faster, less
dense but more powerful wind sweeps up the LBV wind over the course of
the final 400 kyrs of the star's life. The structure formed is presented in 
the next section.

In Fig.~\ref{wnd-60M} we show the logarithm of density on all three
planes through the location of the 60\,M$_\odot$ star at various times 
through the star's lifespan. After only 1.5\,Myrs, the impact on the 
molecular cloud is already almost equivalent to the total impact of the 40\,M$_\odot$ 
star. The wind has carved channels to allow the flow of wind material
out of the cloud and is now breaking out into the surrounding medium.
After 3.3\,Myrs, at the end of the Main Sequence evolution, the structure
is more extreme, with several deep channels carved through the cloud
and considerable amounts of wind material flowing out of the 
cloud. The reverse shock is clear and very close to spherical, almost
isolated from the parent cloud, at a radius of approximately 20\,pc
from the star. Many clumps towards the lower left of the star have been
ablated and are mass-loading the wind flowing away from the star.
However, towards the upper left of the star, many of the clumps are
still distinct, intact and still comparatively close to the star, especially
compared to the now close-to empty region towards the lower left of the
x-plane. Clearly the distribution of clumps and the position of the star
play a key role in defining the wind-blown structure around the star.

After 2.74\,Myrs, the star enters the LBV phase and again the mass-loss rate increases
and terminal wind speed reduces. As we saw also for the 40\,M$_\odot$
scenario, this slow dense wind forms a high density 
environment around the location of the star. We show the nature of the 
environment after this LBV phase along the third row of Fig.~\ref{wnd-60M}.
After this, during the WR phase of stellar evolution, the variable, faster, 
less dense but more powerful wind is able to sweep the LBV and Main Sequence material
out into the cloud creating the final environment into which the supernova
will go off. By this time, the reverse shock is now up to 35\,pc from the star
and there are no traces of the high-density environment formed during the
LBV phase. The large reverse shock structure dominates the cloud, outside
which multiple channels will allow SN material to be
transported rapidly out of the cloud. We note that some clumps
have again survived this more extreme stellar evolution and remain
relatively close to the star.

In Fig.~\ref{wnd-120M} we show the logarithm of density on all three
planes through the location of the 120\,M$_\odot$ star at various
times through the star's lifespan. The 120\,M$_\odot$ star evolves
through the Main Sequence period in 2.24\,Myrs. The first two rows of
Fig.~\ref{wnd-120M} show the extreme effect this star has on its
environment. By the end of the Main Sequence, half the cloud has been
blown away, as shown in the $x$-plane. The reverse shock, previously
relatively small, now approaches half the radius of the original
cloud.  It is almost spherical - the molecular cloud material has been
carried out by the force of the stellar wind. In the $y$- and
$z$-planes, channels through the cloud are clear. In the $z$-plane,
some molecular cloud material is still reasonably close to the star,
just outside the reverse shock.  During the following LBV phase,
multiple non-spherical shells form around the star, as shown in the
third row of Fig~\ref{wnd-120M}, caused by the oscillatory changes in
the mass-loss rate and wind speed (see Fig~\ref{120evolution}). The
structure of the cloud is quite different depending upon which plane
is considered. Low-density empty voids exist to the left in the
$x$-plane and clumpy structure surrounds the shells formed during the
LBV phase in the $y$-plane.

After the LBV phase, the variable, faster, less dense WR wind is able to 
evacuate the interior of the cloud in all directions, generating a strong
reverse shock that in the empty void extends to more than 50\,pc from
the star. We show the final structure of the cloud at the end of the star's
life, after 3.01\,Myrs, across the fourth row of Fig.~\ref{wnd-120M}. Much
of the cloud has been swept away. Compared to the previous cases, the 
least amount of clumpy cloud material remains around the star - something
we will study in more detail in the following Analysis section. 

In Fig.~\ref{wnd-proj} we show column density, naive emission (calculated
from the radiative cooling energy source term) and 3D visualisations of the
cloud-wind structure at the end of the life of the three most massive stars. In the
40\,M$_\odot$ star case, the blown-out side of the cloud is very clear in the
column density plot. The radiative cooling plot highlights the `emission' from
the isolated LBV/WR shell very clearly, as discussed in more detail in the next
sub-section. In the case of the 60\,M$_\odot$ star, the column density
projected along the $y$-axis highlights the fact that whilst the wind has blown
out a larger section of the cloud than in the previous case (as shown in
Fig.~\ref{wnd-60M}), the line-of-sight has strong effect on how obvious this
blow-out is to the observer. The naive emission from the radiative cooling
energy source term highlights the much greater extent of the more 
powerful WR wind in this case, reaching throughout the cloud. Column 
density shown collapsed along the $z$-direction for the 120\,M$_\odot$ star
does show very clearly how much of the cloud has been dispersed from the 
original structure, as compared to how much is left in the top half of the cloud.
Distinct in this plot and the plot for the 40\,M$_\odot$ star are the radially aligned
`spokes' of the cloud after sculpting by the stellar wind. 
The hot, fast WR wind has triggered 'emission' all over the structure in the
case of the 120\,M$_\odot$ star. Of course the `emission' assumes optically thin
conditions and true estimates of emission require radiative transfer to be 
simulated alongside the hydrodynamic evolution of the cloud-wind interaction.
In the third column of Fig.~\ref{wnd-proj} we show a complex 3D isosurface
of the wind scalar on half the volume (indicated by the colour scale for the 
domain x$\leq$0) and a second isosurface of the high density clumpy cloud 
structure (indicated by the black isosurfaces at
$\rho$=25\,cm$^{-3}$). In all three star cases, the yellow core highlights the
inner core of the wind-blown bubble as well as the increasing extent of the
wind-carved channels with greater stellar mass, reaching off the domain in 
many places for the case of the 120\,M$_\odot$ star. Noticeable also from 
these 3D visualisations is the widening distribution of the high density structure
(indicated by the black clumps), dispersed by the increasingly powerful stellar 
winds with increasing stellar mass.

\subsection{The early supernova phase}\label{snphase}

\begin{figure*}
\centering
\includegraphics[width=165mm]{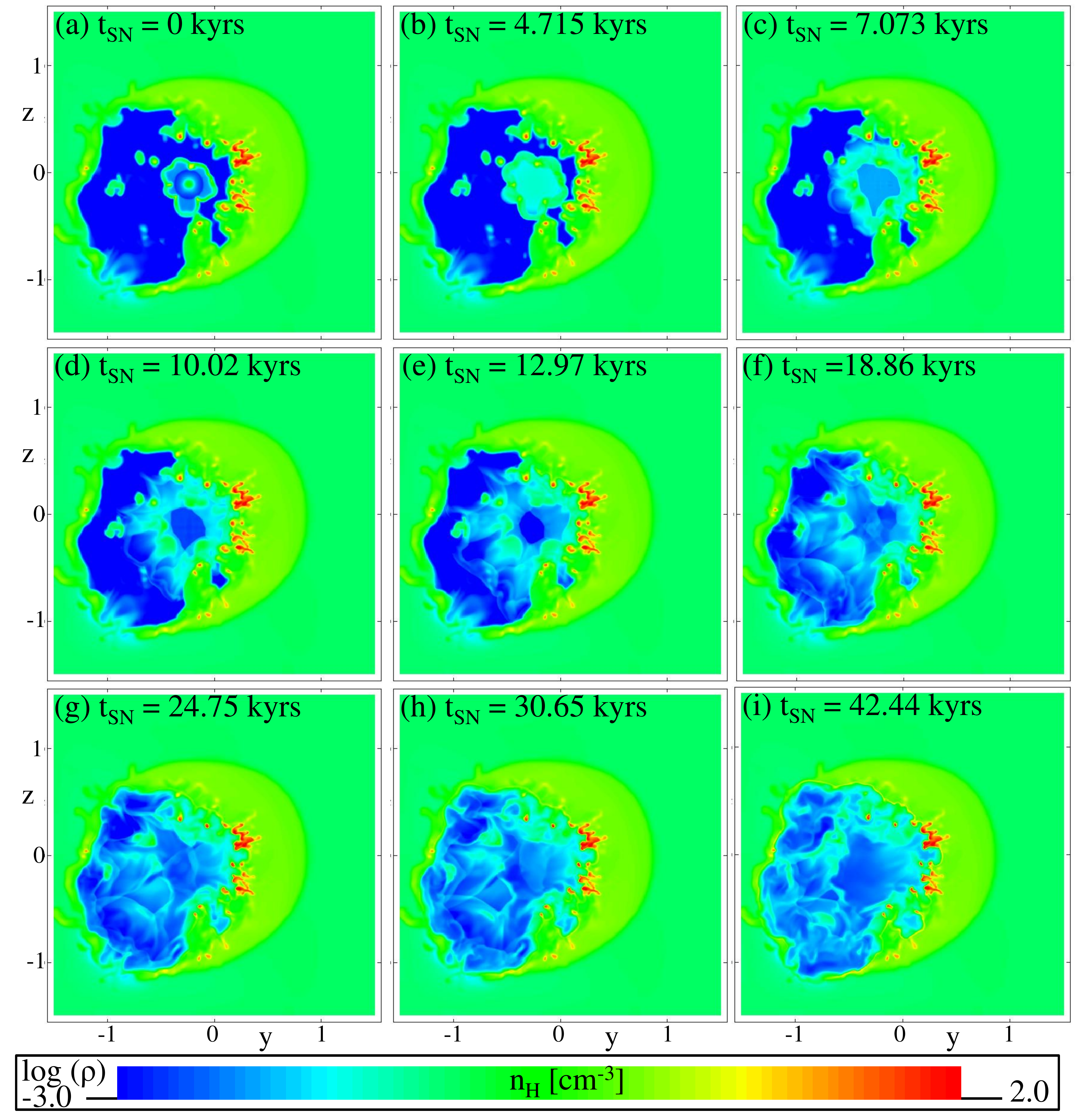}
\caption{Early SN-cloud-wind interaction for a 40\,M$_\odot$ star. 
Shown is the logarithm of mass density at various times on the 
plane at $x$ = -0.025. Length is scaled in units of 50\,pc.
Raw data: doi.org/10.5518/201.} 
\label{sn-40M}
\end{figure*}

\begin{figure*}
\centering
\includegraphics[width=165mm]{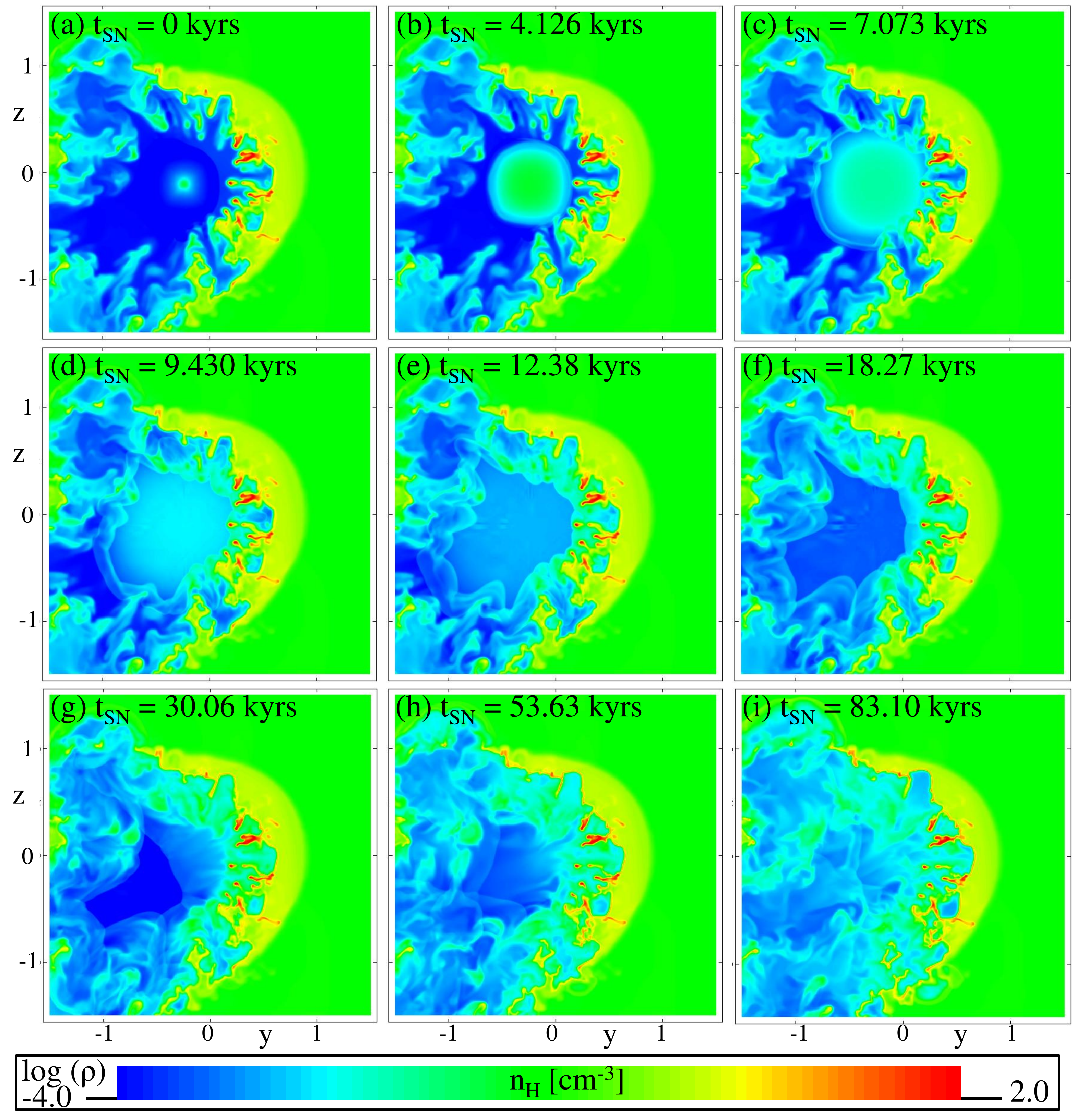}
\caption{Early SN-cloud-wind interaction for a 60\,M$_\odot$ star. 
Shown is the logarithm of mass density at various times on the 
plane at $x$ = -0.025. Length is scaled in units of 50\,pc.
Raw data: doi.org/10.5518/201.} 
\label{sn-60M}
\end{figure*}

\begin{figure*}
\centering
\includegraphics[width=165mm]{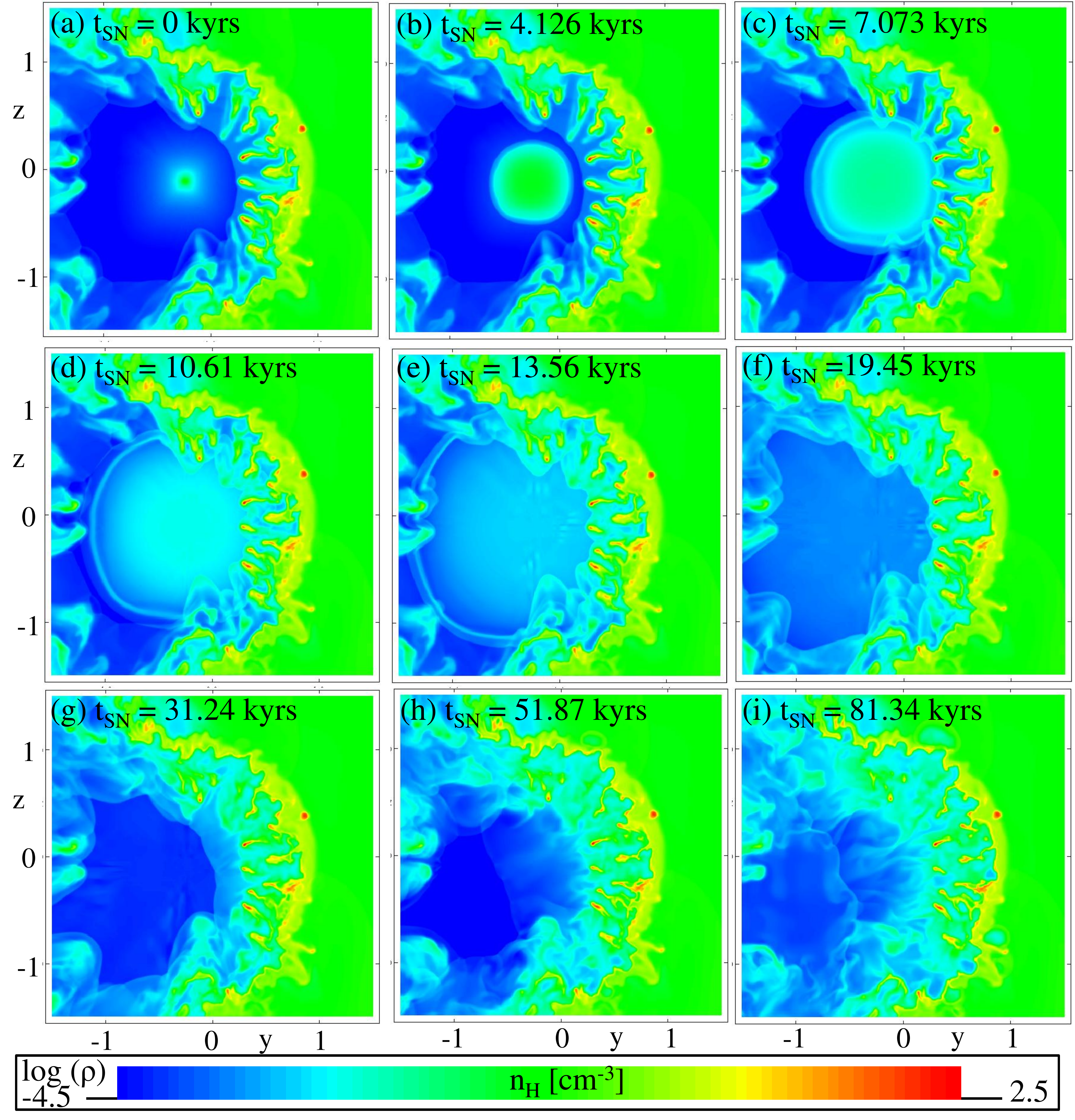}
\caption{Early SN-cloud-wind interaction for a 120\,M$_\odot$ star. 
Shown is the logarithm of mass density at various times on the 
plane at $x$ = -0.025. Length is scaled in units of 50\,pc.
Raw data: doi.org/10.5518/201.} 
\label{sn-120M}
\end{figure*}

The environment into which the SN mass and energy from the
40\,M$_\odot$ star are injected is shown in Fig.~\ref{sn-40M}(a).  The
non-spherical high density shell structure formed during the WR stage
of evolution is centred on the location of the star, distinct from,
but strongly influenced by, the surrounding molecular cloud structure.
In 4700\,yrs the supernova remnant (SNR) has propagated far enough to
fill the LBV/WR shell and after 7073 years the SNR has overrun the
shell and is now expanding into the star's wind-blown bubble, as shown
in panel (b) and (c) of Fig.~\ref{sn-40M}. After 10,000 yrs, shown in
panel (d), the forward shock of the SNR can clearly be seen expanding
out into the bubble. Meanwhile, internal shocks reflecting off the inner shell
are refilling the SNR. Clump remnants that had survived the wind phase
are compressed by the SNR over the next 10,000 years (panels (e) and
(f)) and then accelerated outwards, towards the edge of the wind
bubble. By 42,000 years, in the final panel of Fig.~\ref{sn-40M},
the internal shocks running back towards the location of the star from the 
edge of the wind bubble are acting to generate even more complex
structure inside the SNR. The SNR now continues to expand, encompassing
the wind bubble. The forward shock of the SNR reaches the edge of the
grid after 69,000 years, at the point where the SNR is closest to the edge 
of the grid in this panel.

In our previous work, studying the evolution of 40\,M$_\odot$ star in
a sheet-like cloud, we found the stellar wind generated a tunnel
through the cloud, through which the SNR was then able to escape the
parent cloud in a matter of only 30,000 years. Here, the SNR is able
to progress through the wind-blown bubble equally as quickly, going
beyond the extent of the original parent cloud where the wind-blown
bubble had swept up and then beyond the molecular cloud itself.  In
other directions though, the SNR has been rapidly deccelerated as it
enters the parent molecular cloud - only in panel (e) of
Fig.~\ref{sn-40M}, 13,000 years after the SN event, do we begin to
clearly see the SNR progressing into clumpy cloud material. So in this
case, as in the magnetically-influenced case of the 40\,M$_\odot$ star
in Paper II, the wind-blown environment is the key to the evolution of
the SNR. Progress in the hydrodynamic case is dramatically slowed by
the denser clumps and the inter-clump material, as compared to
progress through the low-density wind cavity. We now examine the same
intra-cavity phase for the 60\,M$_\odot$ and 120\,M$_\odot$ stars.

In Fig.~\ref{sn-60M} we show the logarithm of density on planes
through the location of the 60\,M$_\odot$ star at times after the SN
event corresponding approximately to those shown in
Fig.~\ref{sn-40M}. Immediately clear in panel (a) is the larger size
of the wind-blown bubble and also the lack of any LBV/WR shell. The
late-stage wind has blown out through the entire cavity formed in the
preceding phase. The reverse shock of the WR wind is visible above and
to the right of the star's location, located at a stand-off against
the remaining clumpy cloud material. In the other directions, the
shock of the WR wind is not visible in this figure due to the scaling
chosen to show the supernova evolution, but can be seen in the final
row of Fig.~\ref{wnd-60M}.  The supernova explodes into this
environment. For the first 4000 years the SNR expands into WR wind,
unhindered by material from the preceding phases, or cloud material,
as shown in panel (b).  By 7000 years the SNR has passed the reverse
shock of the WR wind in all directions and is now interacting with the
remaining cloud material close to the star. Shocks are beginning to
propagate down the channels between the network of clumps, whilst
stand-off bow shocks are also forming ahead of the radial spokes of
the network. Meanwhile, in the opposite direction, the SNR continues
to expand into shocked LBV/WR wind material. This scenario of free
expansion in one direction with interaction in the other now describes
the next ~20,000 years of SNR evolution as shown in the middle row of
Fig.~\ref{sn-60M}. Compared to the case of the 40\,M$_\odot$ star,
the wind structure is much larger and the channels off the grid have
allowed the forward shock to leave the grid much earlier, specifically 
only 15,000 years after the SN event, propagating down the widest
wind-blown channel shown on the left of these panels. By 50,000 years,
the interactions of the SNR with this structure have setup multiple
bow shocks which interact and form an even more complex
structure. Channels allowing the early propagation of the SNR off the
grid have now been filled with multiple wide bow-shock structures, as
has the inside of the remnant. It is of particular interest here to
note the survival of the spoke-like network of clumpy material to the
left of the star.  As the preceding wind rearranged the individual
clumps into aligned spokes, the cold cloud material aligned in the
spokes is self-shielded against the passage of the SNR (and the
preceding WR wind). Thus, whilst the head of each spoke is ablated,
the length of the structures at 10-20\,pc means much of the structure
survives beyond this early stage of the SNR - clearly until at least
80,000 years beyond the SN event, as shown in panel (i). We will
explore the further evolution of these structures in the next
sub-section and analyse the amount of cold cloud component remaining
in the following Section.  Some slight numerical artifacts can be seen
in the interior of the SNR at times, caused by derefinements of the AMR 
scheme in the smooth but ballistic and expanding flow. These have no 
effect on the passage of strong internal shocks, reflected off the
cloud-wind structure, as can be seen across the third row of the figure.

In Fig.~\ref{sn-120M} we show the logarithm of density on planes
through the location of the 120\,M$_\odot$ star at times after the SN
event corresponding approximately to those shown in Figs.~\ref{sn-60M}
and ~\ref{sn-120M}.  Clear again in panel (a) is the greater effect
upon the cloud of the stellar wind - half the cloud has been blown
away. Again there is no LBV/WR shell - the reverse shock of the WR
wind is visible to the right of the star's location, located at a
greater stand-off distance against the remaining clumpy cloud material
than in the case of the 60\,M$_\odot$ star. In the other directions,
the reverse shock of the WR wind is almost off the grid and far
outside the original extent of the parent cloud
(r$_{cloud}\approx1$). Panel (b) shows the SNR expanding into the
undisturbed WR wind cavity for the first 5,000 years or so. After this
time, the SNR passes the WR reverse shock boundary, creating internal
structure at the edge of the SNR. It also begins to interact with the
remaining `spokes' of cold cloud material, ablating the inner end of
the radially-aligned spoke structure.

Over the next 10,000 years (second row of Fig~\ref{sn-120M}), the
forward shock of the SNR continues to expand into shocked WR wind
material in one direction, whilst interacting with remaining cloud
material in the other. This makes it easy to see how the SNR evolves
differently under almost unhindered expansion versus expansion into
multiple cold molecular clumps. This also highlights, as with the
previous cases, that it is possible to observe {\it both} scenarios in
one supernova event. The forward shock leaves the grid after 
approximately 15,000 years, similarly to the 60\,M$_\odot$ star case.
Across the third row of Fig.~\ref{sn-120M} it is
possible to see the same SNR interaction behaviour as in the previous
case of the 40\,M$_\odot$ star, in that the SNR expands along the free
channels, only in this case the left side of the cloud is so dispersed
that it is more like remnants of cloud material in a low-density
shocked wind medium, than channels through shocked medium.  For this
reason the bow-shocks forming ahead of the remaining material against
the SNR are not able to intersect to the extent that a reverse shock
propagates back towards the explosion site. Indeed, the evolution of
the bow-shocks over time shows the relatively rapid cooling of these
structures, as they fall back onto the clumps they were initially
detached from.  A large void in the centre of the SNR now
exists. Evolving rapidly into this void, launched off the remaining
cloud material to the right, is an internal shock. This can clearly be
seen moving across the grid with a negative $y$-velocity in the bottom
row of Fig.~\ref{sn-120M}. By 50,000 years, this structure has passed
the original location of the SN event. This creates the
observationally interesting scenario of an offset `centre' of the SNR,
compared to the original location of the star (and the current
location of any remaining stellar remnant). This particular simulation
demonstrates the case of the stellar remnant being close to the edge
of the SNR within a comparatively short 50,000 years since the SN
event.

\subsection{The late supernova phase}\label{latesnphase}

\begin{figure*}
\centering
\includegraphics[width=165mm]{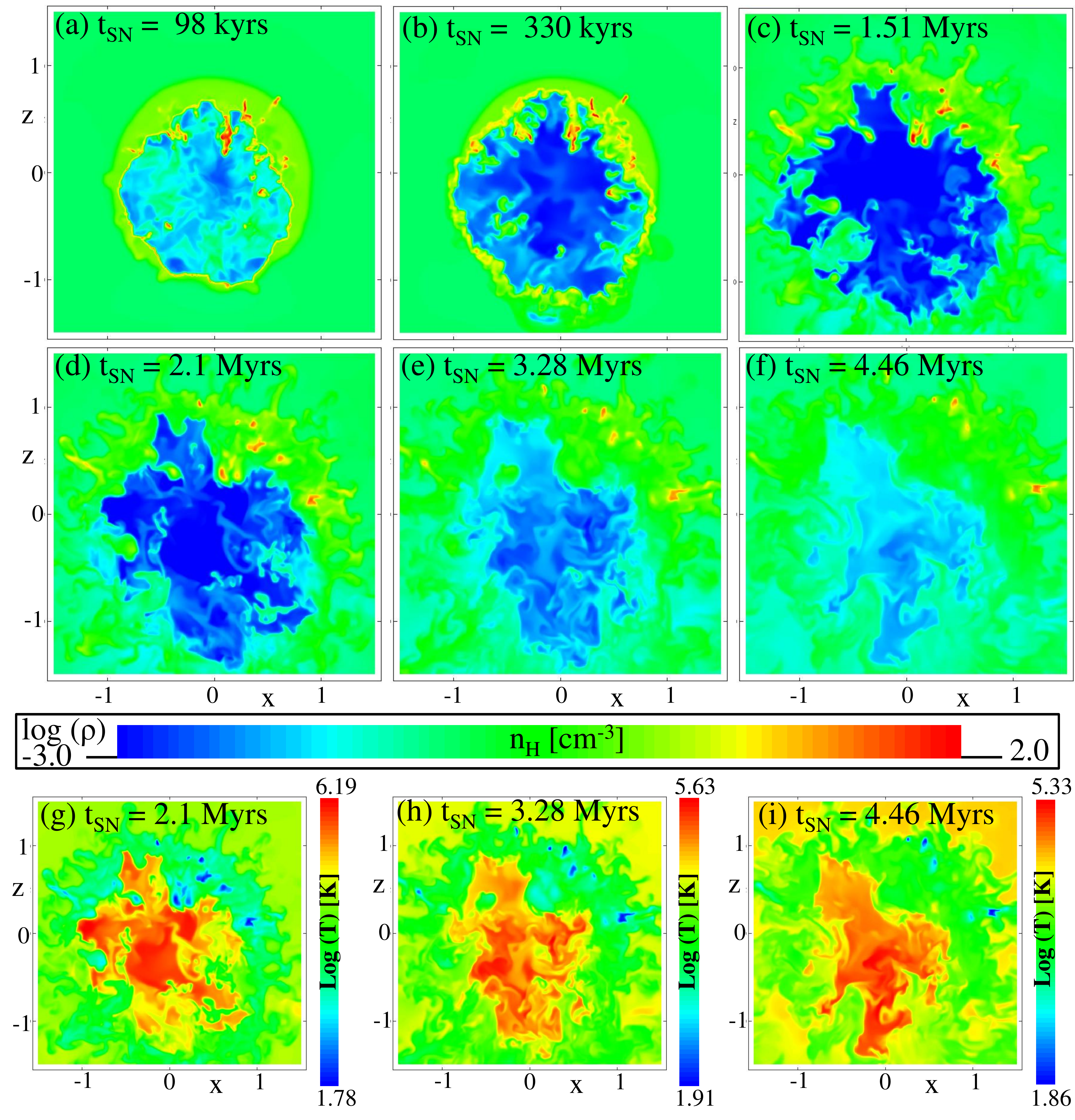}
\caption{Late SN-cloud-wind interaction for a 40\,M$_\odot$ star. 
Shown is the logarithm of mass density and temperature at various 
times on the plane at $y$ = -0.2351. Length is scaled in units of 50\,pc.
Raw data: doi.org/10.5518/201.} 
\label{snlate-40M}
\end{figure*}

\begin{figure*}
\centering
\includegraphics[width=160mm]{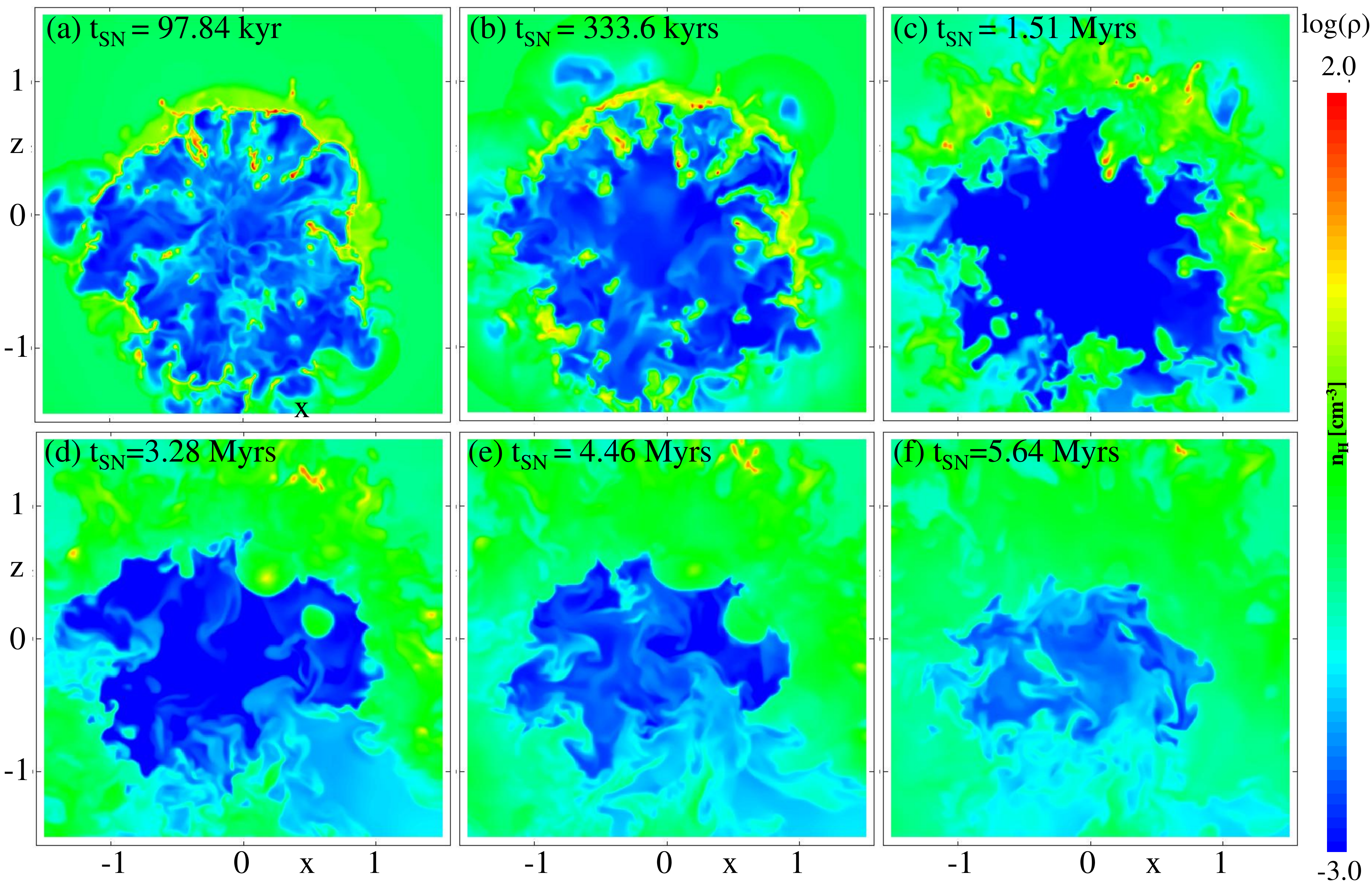}
\caption{Late SN-cloud-wind interaction for a 60\,M$_\odot$ star. 
Shown is the logarithm of mass density at various 
times on the plane at $y$ = -0.2351. Length is scaled in units of 50\,pc.
Raw data: doi.org/10.5518/201.} 
\label{snlate-60M}
\end{figure*}

\begin{figure*}
\centering
\includegraphics[width=160mm]{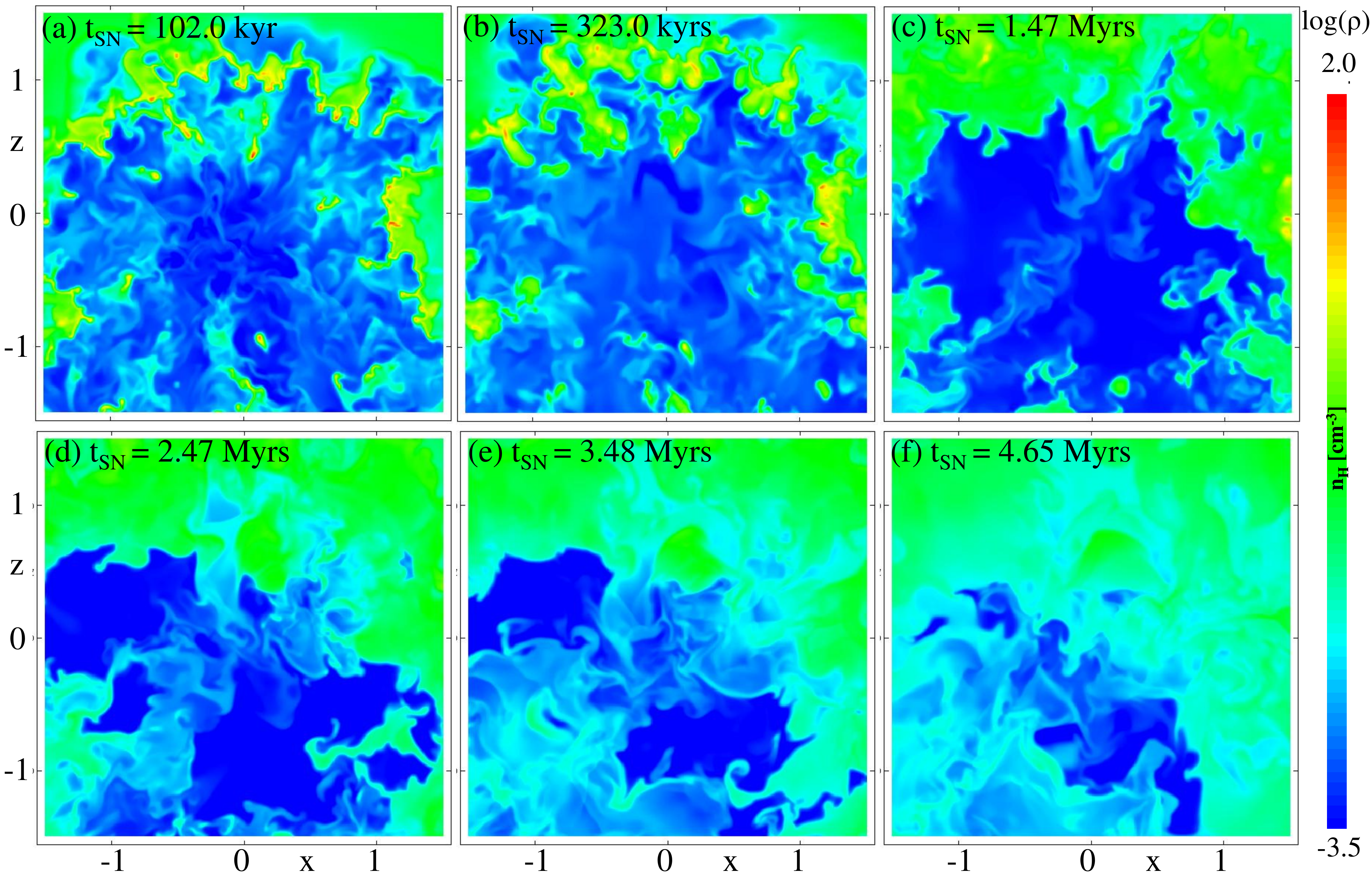}
\caption{Late SN-cloud-wind interaction for a 120\,M$_\odot$ star. 
Shown is the logarithm of mass density at various 
times on the plane at $y$ = -0.2351. Length is scaled in units of 50\,pc.
Raw data: doi.org/10.5518/201.} 
\label{snlate-120M}
\end{figure*}

In Fig.~\ref{snlate-40M} we show the late post-SN evolution of the
40\,M$_\odot$ star simulation on $y$-planes through the location of
the star. We show $y$-planes in this figure (as opposed to the
$x$-planes shown for the early SN stage), as the SNR leaves the
computational volume later on in the $y$-plane (due to the position of
the star and structure of the cloud) and hence boundary effects come
into play later on this plane. After 98,000 years, as shown in panel
(a), the SNR can be seen escaping the molecular cloud, having now
swept up the wind material into the thin shell, punctured by the
spokes of remaining cloud material. Over the next few hundred thousand
years, internal shocks bounce back from the edge of the SNR, and the
ablation of the remaining cloud material increases due to the passage
of these shocks within the internal void of the SNR. The SNR has also
expanded and cooled far enough to reach the radiative phase of
evolution. In panel (b), the forward shock is becoming subject to
Rayleigh-Taylor instabilities and beginning to break up into
individual clumps, in the same way as noted previously in Paper II. 
The remnants of the molecular cloud survive this
phase reasonably well. 

By 1.51\,Myrs, the forward shock of the SNR has reached the edge of
the computational volume and the simulation after this point should
not be taken too literally, since boundary effects have come into
play.  Nevertheless, the later evolution can be used as a guide to the
sort of behaviour one might expect to see. We are now particularly
interested in the fate of the remaining cloud material. Considering
the second row of panels, (d) at 2.1\,Myrs, (e) at 3.28\,Myrs and (f)
at 4.46\,Myrs post-SN, it is clear that the molecular cloud has been
dispersed as all that is left are a few disparate high-density clumps.
It's clear by looking at the temperature plots at equivalent times, as
shown in the third row of the figure, that these are still cold
($\leq$100\,K), even after the passage of the SNR and are the final
remaining parts of the molecular cloud. The structure of the cloud is
destroyed, but parts of the cold cloud material remain, much as
occurred in the case of the 40\,M$_\odot$ star in a
magnetically-collimated sheet-like cloud in Paper II. As also noted in
Paper II, self-gravity in the dispersed clumps is now likely to
dominate their evolution, possibly leading to further star formation.
It is highly likely that the disruption of the molecular cloud is
caused by the SN event and not influenced by any numerical effects, as
the outward motion of the clumps can be traced back to before the SNR
reached the edge of the computational volume. We will consider in the
following cases whether this can be confirmed further, by examining
the rate of cloud destruction in each case, but to be more confident
of this result, future computationally-costly simulations with a
larger volume are necessary. Further questions over the fate of the
remaining molecular material and whether the SN triggers any further
formation of cold material by disrupting the thermal stability of the
wind/cloud material are addressed in Section \ref{analysis}. Some
indications are given by the decreasing lower limit of the temperature
range going from panels (h) to (i), indicating the presence of
ever-colder material on this plane.  Note also the `turbulent' nature
of the interior of the SNR, caused by the interaction of multiple
shocks. Even after 2\,Myrs, the temperature inside the SNR is still
above 10$^6$\,K. The SNR cools noticeably over the next
2.5\,Myrs. With a cooling and dispersing remnant, gravity plays an
increasingly important role. We have simulated on to 6\,Myrs post-SN,
but are unable to confidently assess the balance in the simulation
between the evolution caused by the SNR, by gravity and by the
influence of any boundary effects.

In Fig.~\ref{snlate-60M} we show the late post-SN evolution of the
60\,M$_\odot$ star simulation. Whilst structural similarities to the
40\,M$_\odot$ case are apparent, it should be noted that the SNR is
larger and has left the cloud at a much earlier time.  Clearly the
major reason is the enhanced disruption of the cloud by the wind
during the preceding evolution. While the SNR transitions into the
radiative phase at around the same time, comparing panel (b) in both
this and Fig.\ref{snlate-40M}, it is clear that the rest of the
remnant is much more dispersed.  The same effects come into play as
the star evolves into Myrs post-SN, but in this case it should be
noted that the clumpy remains of the molecular cloud are fewer and
further out from the star. We show the SNR at 5.64\,Myrs post-SN to
display the final turbulent state of the SNR. We analyse the evolution
of the cold molecular material in the computational volume as a whole
in the next Section.

In Fig.~\ref{snlate-120M} we show the late post-SN evolution of the
120\,M$_\odot$ star simulation. The effect of the preceding wind phase
- to allow the SNR to have completely blown through the wind-cloud
structure in the $y$-plane after only 100,000 years - is shown in
panel (a). A slight ``echo'' of the edge of the cloud remains, but
otherwise turbulent conditions dominate the interior of the SNR. This
arc-like echo of the cloud breaks up after 300,000 years of SNR
evolution. After 1.5\,Myrs, the last few high-density cloud components
are leaving the grid. After this, a low density SNR void is surrounded
by average ISM conditions (densities ranging from 10$^{-2}$ to 10$^0$
particles per cm$^3$). There are no cloud components left on this
plane - the cloud has been completely dispersed by the SNR. 

\section{Analysis}\label{analysis}

\subsection{Energy}

\begin{figure*}
\centering
\includegraphics[width=170mm]{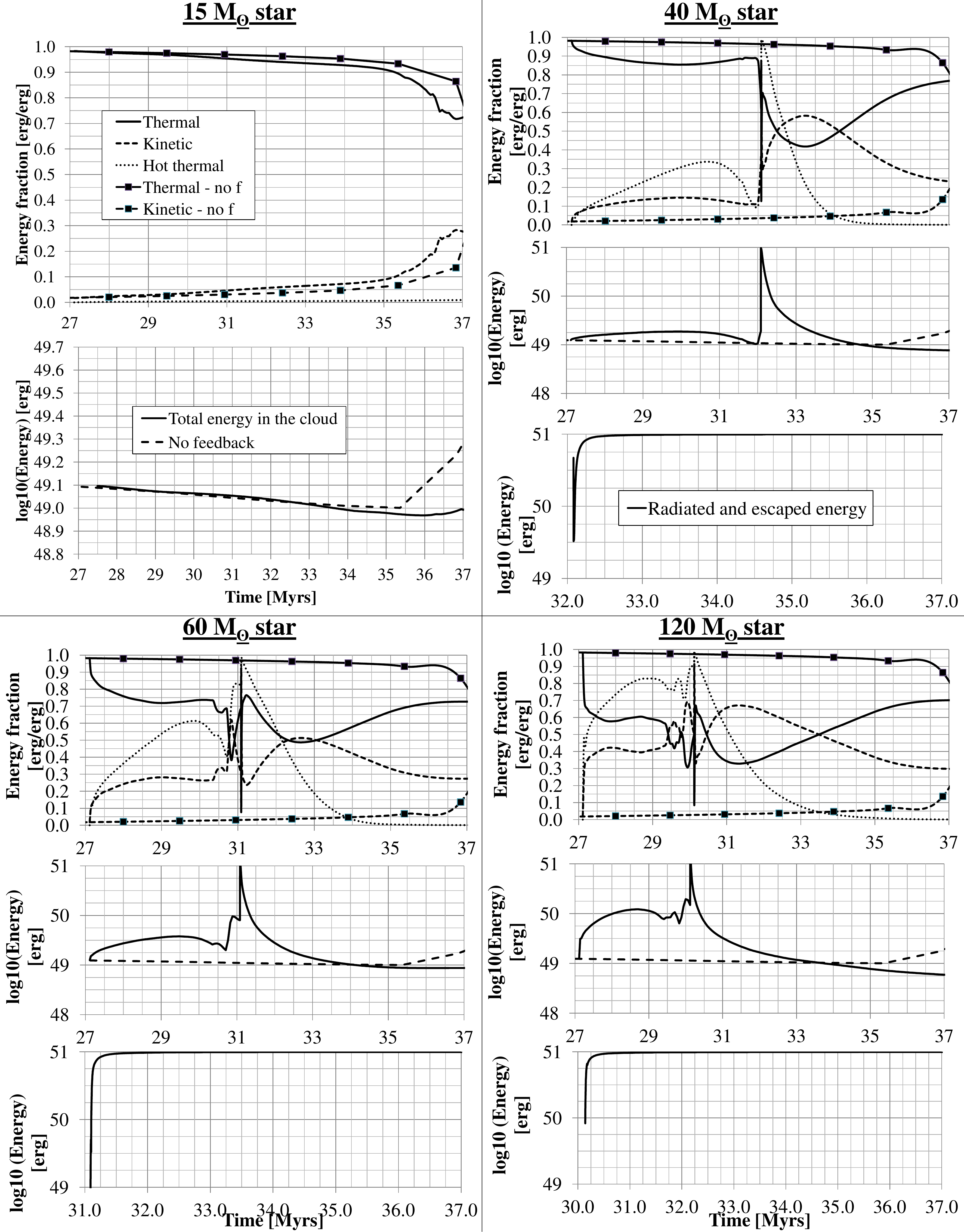}
\caption{Time profiles of energy fractions, totals and escaped amounts
  of energy in the cloud, wind and SN material combined. The thermal and kinetic energy fractions together
  sum to 1.0 i.e. we are ignoring the change in gravitational energy. The 
  `hot thermal' profile indicates the fraction of the thermal energy which 
  is hot (above $10^{4}$\,K), and so also has a range between 0.0 and 
  1.0. Lines with markers indicate the energy behaviour in the cloud with no
  stellar feedback. On the third row of the 40, 60 and 120 M$_{\odot}$ 
  star plots, the ``radiated and escaped energy" is the total energy 
  subtracted from the injected energy of the SN, 10$^{51}$ erg. Note
  the different timescales in these plots. Raw data: doi.org/10.5518/201.}
\label{energy}
\end{figure*}

In Fig.~\ref{energy} we show how the distribution of energy between
kinetic, thermal and hot thermal (i.e. material above 10,000K) in the
cloud, wind and SN material combined, varies
over time following the introduction of stellar feedback, with
reference to a simulation of the cloud evolution without feedback for
each star. The thermal and kinetic energy fractions together sum to
1.0, i.e. we are ignoring gravitational energy in this figure. Note
that the hot thermal profile indicates the fraction of the thermal
energy which is ``hot''. In the second plot for each star, we show how
the total energy in the cloud varies over time, with reference to the
simulations of the cloud without feedback. The third plot for the 40,
60 and 120\,M$_\odot$ star cases shows the post-SN ``radiated and
escaped energy", i.e. energy that is radiated away or leaves the grid
dynamically - we do not differentiate between these two means of
energy loss in the plot.

\begin{figure*}
\centering
\includegraphics[width=135mm]{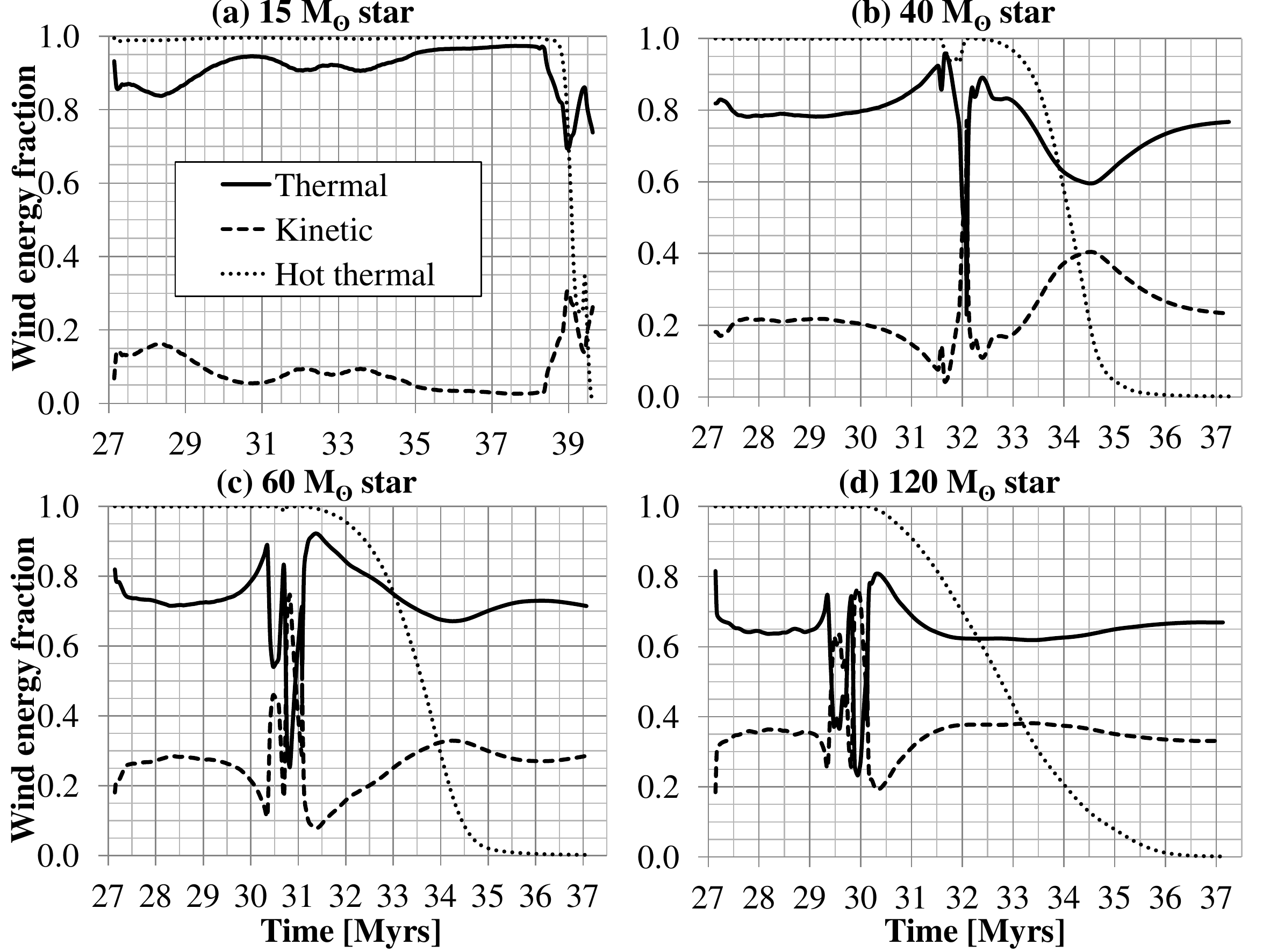}
\caption{Time profiles of the energy fractions of the stellar wind
  material as traced by the advected scalar $\alpha_{wind}$. The thermal 
  and kinetic energy fractions together sum to 1.0 i.e. we are ignoring the 
  change in gravitational energy. The 
  `hot thermal' profile indicates the fraction of the thermal energy which 
  is hot (above $10^{4}$\,K), and so also has a range between 0.0 and 
  1.0. Raw data: doi.org/10.5518/201.}
\label{wndenergy}
\end{figure*}

\begin{figure*}
\centering
\includegraphics[width=135mm]{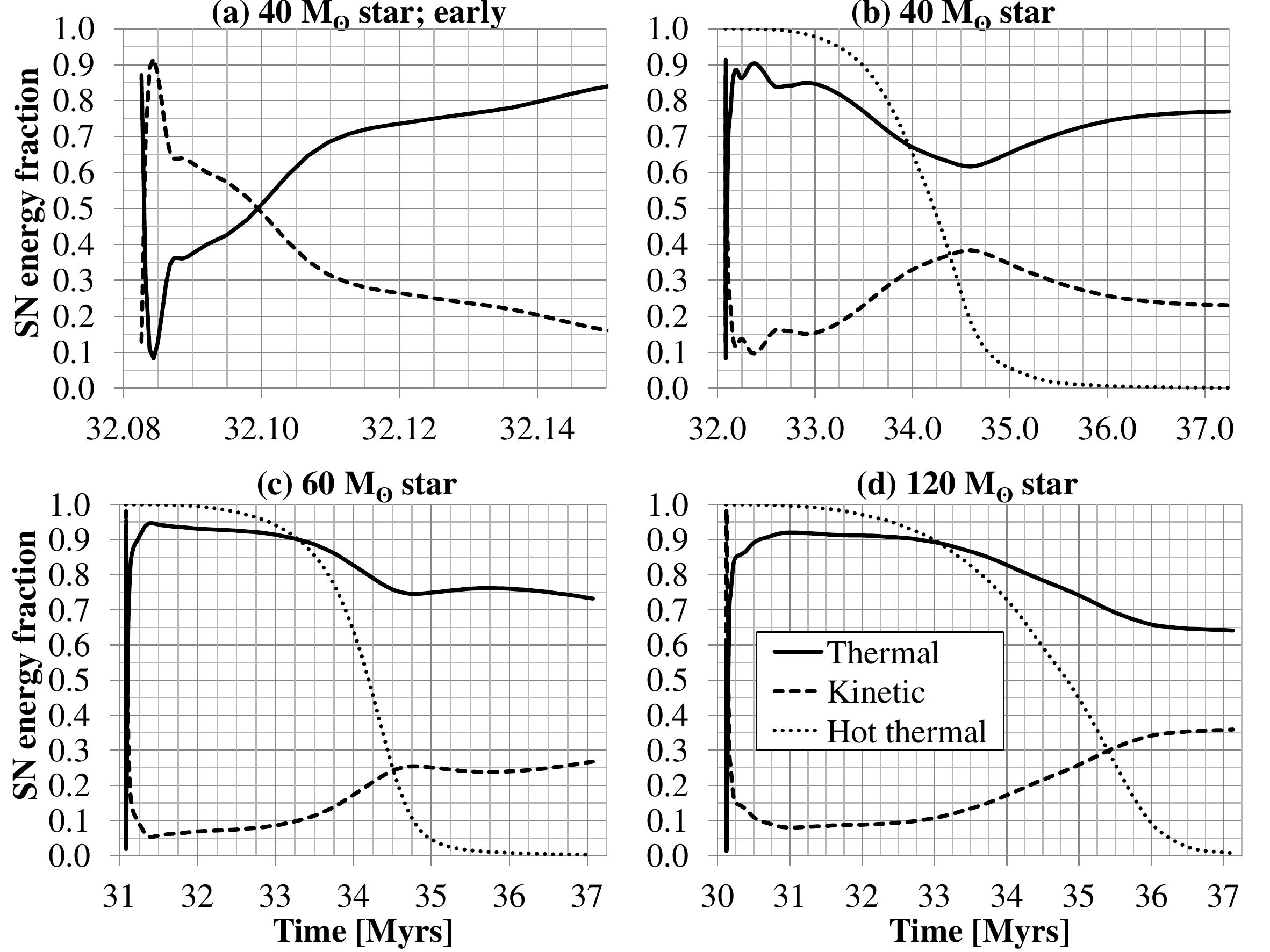}
\caption{Time profiles of the energy fractions of the supernova
  ejecta, as traced by the advected scalar $\alpha_{SN}$The thermal 
  and kinetic energy fractions together sum to 1.0 i.e. we are ignoring the 
  change in gravitational energy. The 
  `hot thermal' profile indicates the fraction of the thermal energy which 
  is hot (above $10^{4}$\,K), and so also has a range between 0.0 and 
  1.0. Raw data: doi.org/10.5518/201.}
\label{snenergy}
\end{figure*}

For the first 10\,Myrs of the 15\,M$_{\odot}$ star case, the energy
fractions do not change considerably during the wind phase, as
compared to the reference case.  This was expected from the minimal
dynamic and structural impact the star has on the cloud. In fact the
stellar wind introduces so little energy, that the overall energy in
the cloud reduces over the first 10\,Myr of the star's life, as
compared to the `no feedback' reference case. This likely arises due
to relatively efficient radiative cooling of the predominantly thermal
energy injected by the stellar wind, perhaps due to weakly compressing
and heating neighbouring cloud material. From 10\,Myr post star
formation (37\,Myrs in the figure), the collapse of the cloud
dominates the evolution. The stellar wind has little effect on this
and after 37\,Myrs, these results contain no meaningful information.

In the 40\,M$_{\odot}$ star case, the stellar wind has a clear effect
on the energy fractions during the wind phase. The stellar wind
supplies and creates large amounts of kinetic and hot thermal energy,
raising both fractions above the close-to-zero reference case.  Given
that the wind blows a hot bubble, it's likely that radiative losses
are responsible for the drop in total energy approaching the end of
the star's life, before the introduction of 10$^{51}$\,erg of thermal
energy and 10\,M$_{\odot}$ of material in the SN event.  The
introduction of this thermalised SN kinetic energy is most obvious in
the total energy plot, as the spike just after 32\,Myrs. Since the SN
energy is injected thermally, both the thermal energy and the `hot
thermal' phase rise sharply at this time. The third plot for the
40\,M$_{\odot}$ star case shows that the SN energy is rapidly
transported and radiated out of the simulation. Note that the forward
shock of the SNR first reaches the edge of the computational volume
approximately 69,000 years after the SN event. At approximately
1\,Myr post-SN, the energy balances are most different from the
reference case without feedback (apart from the first few thousand
years of SN evolution, which we discuss in more detail later in this
sub-section). After this, as the SNR escapes from the cloud/wind
structure into the surroundings and then off
the grid, the kinetic energy fraction falls and the thermal energy
fraction rises. This behaviour continues, albeit at a decreasing rate,
until the end of the simulation. The actual amount of thermal energy
drops until approximately 2.5\,Myrs post-SN and then plateaus, whereas
the kinetic energy simply continues to drop during the simulation
post-SN. The plateauing of thermal energy corresponds to when the hot
thermal fraction of energy remaining on the grid has fallen to close
to zero - much of the material on the grid has now dropped to
temperatures below 10,000K and densities in the range 0.01-1
cm$^{-3}$, with random motions at a range of velocities and scales
(except for the last remaining gas in the hot, central void of the
SNR, but this is at low density). This is fairly characteristic of the
warm, neutral medium, formed in this case from evolution of an
isolated single star, approximately 2.5\,Myrs post-SN.

\begin{figure*}
\centering
\includegraphics[width=170mm]{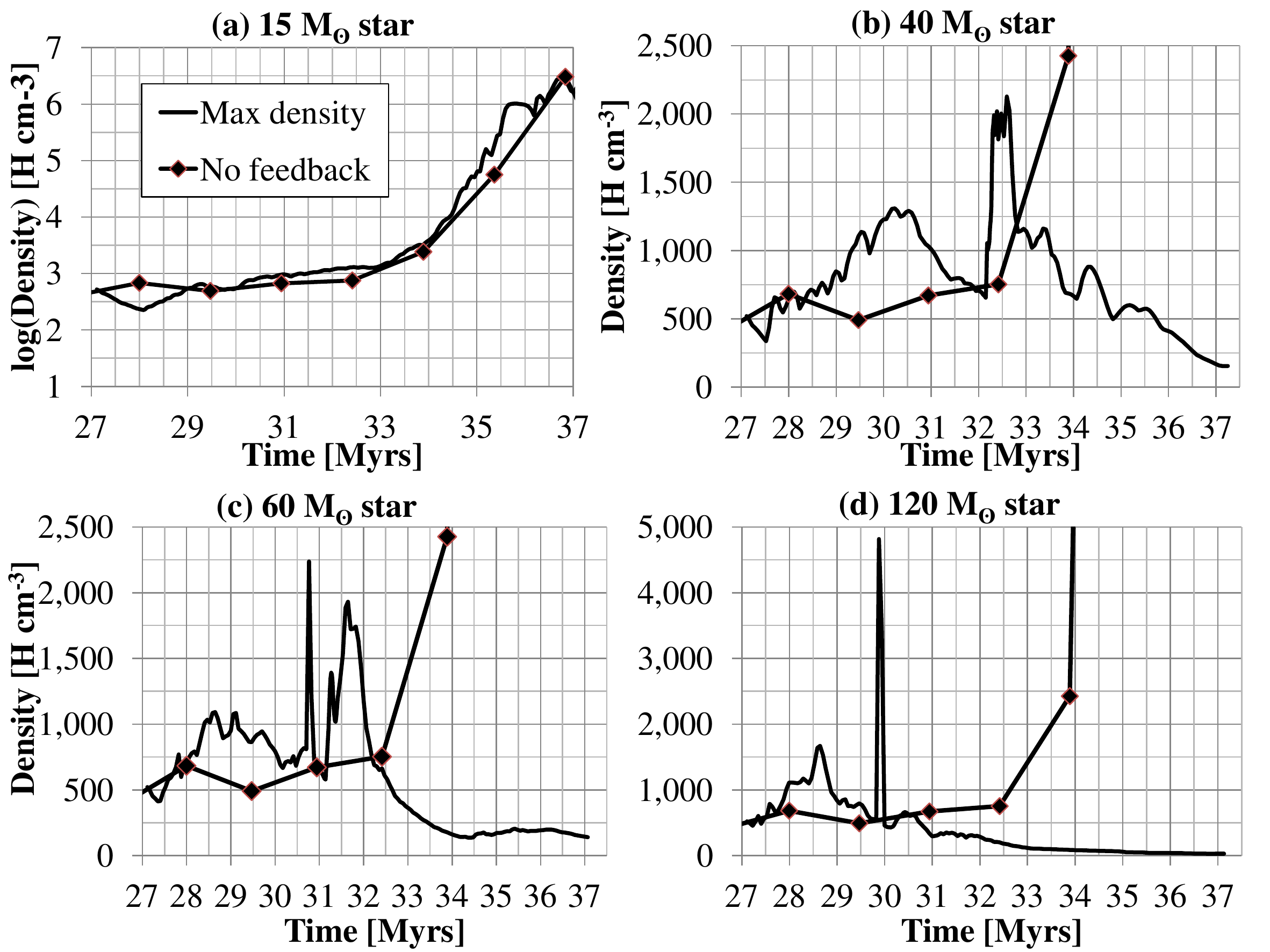}
\caption{Maximum densities reached in the stellar feedback simulations 
(solid lines) shown with the maximum density in a continued simulation 
of the same initial conditions without feedback (lines with markers) for reference.
Note that without feedback, the cloud collapses to unrealistic densities for
the resolution of the simulation. The SN occurs at $t = 32.1$, 31.1
and 30.1\,Myr for the $M = 40$, 60 and 120\,M$_{\odot}$ cases.
Raw data: doi.org/10.5518/201.} 
\label{maxrho}
\end{figure*}

In the 60\,M$_{\odot}$ star case, the wind effect upon the thermal and
kinetic energy fraction is an amplified version of the 40\,M$_{\odot}$
star case. During the Main Sequence wind phase, the amounts of both
thermal and kinetic energy on the grid increase smoothly for
2.5\,Myrs, then decrease smoothly, with the amount of kinetic energy
increasing and decreasing more than thermal energy. In the final
stages of the star's evolution, the total energy in the simulation is
increased by a factor of 5 by the stellar wind, with a sharp increase
in amounts of both thermal and kinetic energy. This corresponds
closely with the increased energy injection rate of the stellar wind,
as shown in Fig.~\ref{60evolution}. The kinetic energy increases by a
factor of 10, leading to the apparent drop in the thermal energy
fraction, although there is no drop in the actual amount of
energy. Just before the supernova, ten times more energy is present in
the simulation than in the reference case, or the 40\,M$_{\odot}$ star
case (which is close to the reference case pre-supernova). One must
bear in mind that a substantial amount of the total wind energy up to
this point has already been transported off the grid, while some has
also been radiated away.

The SN event increases the amount of energy on the grid by a factor of
10. The post-SN evolution of the energy fraction and the total energy
in the simulation evolves in a similar way to the 40\,M$_{\odot}$ star
case, with an initial peak in kinetic energy (and trough in thermal
energy) as the SNR blows out of the cloud. In this case though, the
forward shock leaves the grid along the wind-carved channels much
more quickly, first reaching the edge of the computational volume after
only 15,000 years. The initial peaks are followed by similar
evolution between the thermal, hot thermal and kinetic energy
fractions when compared to the 40\,M$_\odot$ star case.  
The thermal energy plateau of warm, neutral medium is
reached on the same timescale.  Whilst the detailed evolution of the
cloud with feedback from a 60\,M$_{\odot}$ star is visibly very
different to the 40\,M$_{\odot}$ star case, as shown in the previous
Section, the evolution of the energy fraction, total energy and
radiated/escaped energy is not considerably different.

The 120\,M$_{\odot}$ star case is similar to the case of the
60\,M$_{\odot}$ star. The wind introduces considerably more kinetic
energy though, driving shock heating and a greater fraction of hot
thermal energy. The late stages of stellar evolution are equally
complex and track the variations in energy injection rate, but the
post-SN evolution is remarkably similar. Most interestingly, the
supernova from the 120\,M$_{\odot}$ star very rapidly loses thermal
energy post-SN and the kinetic energy peaks at a higher fraction than
both the 40 and 60\,M$_{\odot}$ star cases. Again, the forward shock
leaves the grid in around 15,000 years, along the low density channels
carved by the wind through the cloud and surroundings. Although the post-SN
material in the 120\,M$_{\odot}$ case does briefly plateau in the
warm, neutral phase between 2.5 to 3.5\,Myrs post-SN, the thermal
energy on the grid then continues to decrease. The grid is dominated
by the hot, low-density void of the inner SNR, which is still cooling,
leading to this effect. Any warm, netural medium would be outside the
grid. This highlights the range upon which each star acts, increasing
with the mass of the star (and hence strength of the wind). A larger
computational volume is clearly necessary to capture the
characteristics of the surrounding medium that the 120\,M$_{\odot}$
star creates post-SN.

In Fig.~\ref{wndenergy} we show the evolution of the energy fraction
of stellar wind material for each star. The wind material injected by
the 15\,M$_{\odot}$ star is almost completely dominated by hot thermal
energy, and has very little kinetic energy, until the end of the
star's life (indicating that the wind is ``bottled up''). The wind
material from the 40\,M$_{\odot}$ star has a greater fraction of
kinetic energy, which increases steeply close to the supernova event
as the star enters its WR phase. Post-SN, the energy of the wind
material follows a similar trend to that of the total energy in the
simulation, as shown in the last figure. The same trends are seen in
the 60 and 120 \,M$_{\odot}$ star cases.  In Fig.~\ref{snenergy} we
show the evolution of the energy fractions of the supernova material
for each star. Clearest in panel (a), showing the first few tens of
thousands of years of the SNR evolution, is the initial conversion of
thermal energy into the kinetic energy of expansion of the ejecta
(causing the thermal energy to drop to only 10\% of the total ejecta
energy), followed by a rapid increase to 35\% as the expanding ejecta
passes through a reverse shock and is rethermalized to an 85\%
fraction. Over the next 4\,Myr, the thermal energy fraction drops to
60\% of the total (but since material is lost off the grid one cannot
draw any further conclusions).

\subsection{Maximum density}

In Fig.~\ref{maxrho} we show the maximum densities reached in the
simulations, compared to the maximum density at the same time in the
reference case without stellar feedback. In the 15\,M$_{\odot}$ star
case, the maximum density during the wind phase follows the reference
case. The maximum density is more than 10$^6$ H\,cm$^{-3}$ during the
cloud collapse; realistic in the gravitational collapse of a molecular
cloud, but beyond the accuracy of the resolution used in this
simulation. In the other three cases, the stellar winds are able to
create considerably higher densities than in the reference case during
the wind phase, enough for the possibility of a second generation of
star formation with densities over 10$^3$ H\,cm$^{-3}$. Variations are
not due to the stellar wind - it is relatively steady with smooth
variation over the majority of the star's lifetime - but instead must
be due to the compressional effect the stellar wind has on the cloud,
rearranging the structure and compressing ``spokes''.

By the time of the supernova, the maximum density in all three cases
is back to similar levels to the reference case, showing that whilst
the final wind stages inject a considerable amount of mass, densities
elsewhere reduce as the cause of increased density, compression by the
wind, has reduced. The SN instantly raises the density in the
injection location, but the spikes seen in the plots are instead
associated with compression of surrounding LBV/WR material and then
cloud material in the 40\,M$_{\odot}$ star case, and the cloud
material in the 60 and 120\,M$_{\odot}$ star cases. Later peaks in the
maximum density plots, showing against the overall decreasing trend of
maximum density, may be associated with the formation of new cold
material after SN affected material returns to the thermally unstable
phase. Localised star formation, according to our star formation
rules, would occur in all these cases.  Only in the case of the
60\,M$_{\odot}$ star is a late onset increase in maximum density
apparent (post 34.5\,Myrs). This must be associated with increasing
amounts of cold molecular material, even in the warm, neutral
medium. This is investigated further in the next sub-section.

\subsection{Phases}

\begin{figure*}
\centering
\includegraphics[width=170mm]{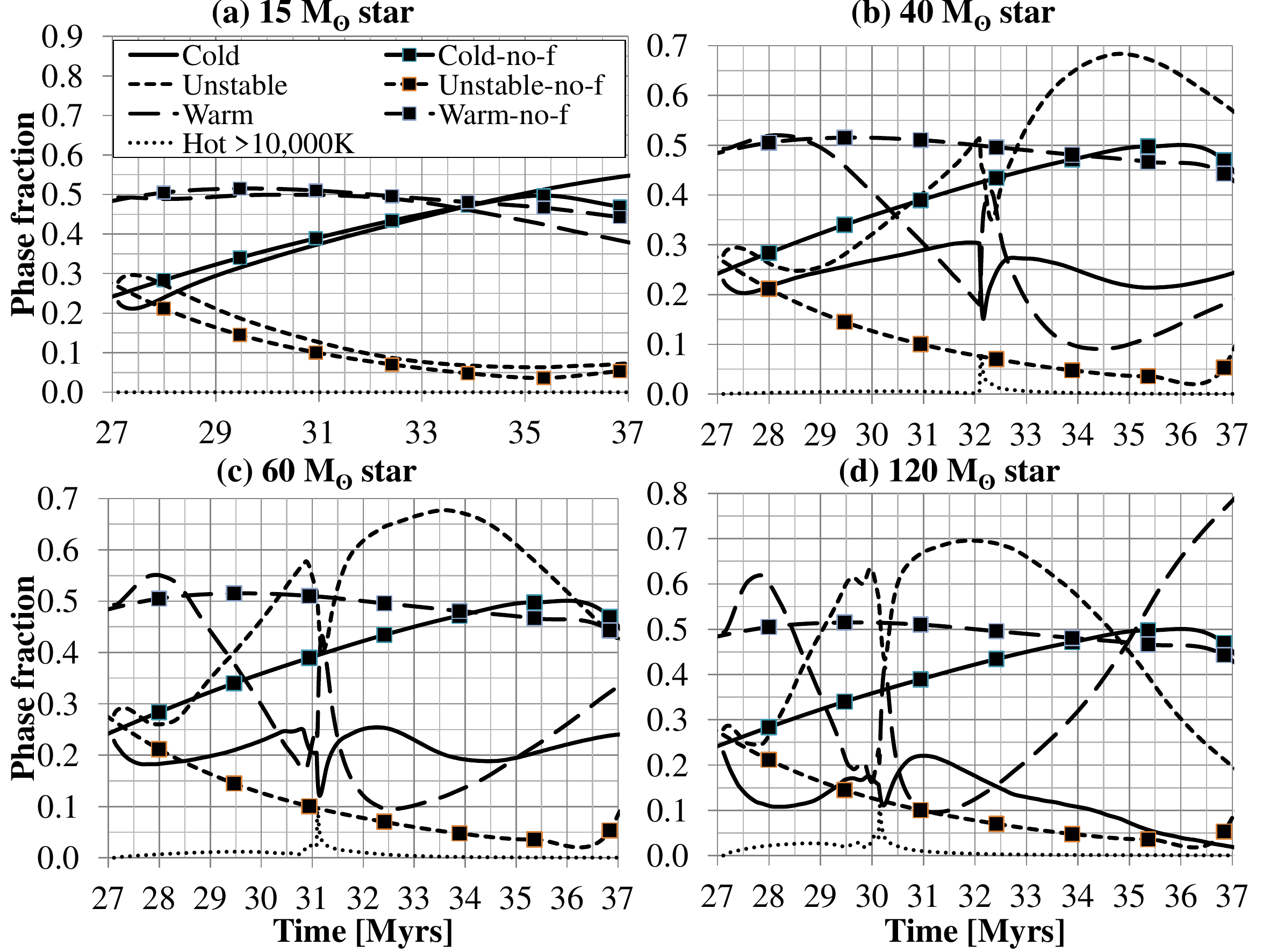}
\caption{Variation of phase fractions with time for the feedback simulations.
The reference case without feedback is indicated by lines with markers in the 
phase fraction plots. Raw data: doi.org/10.5518/201.} 
\label{phases}
\end{figure*}

\begin{figure*}
\centering
\includegraphics[width=150mm]{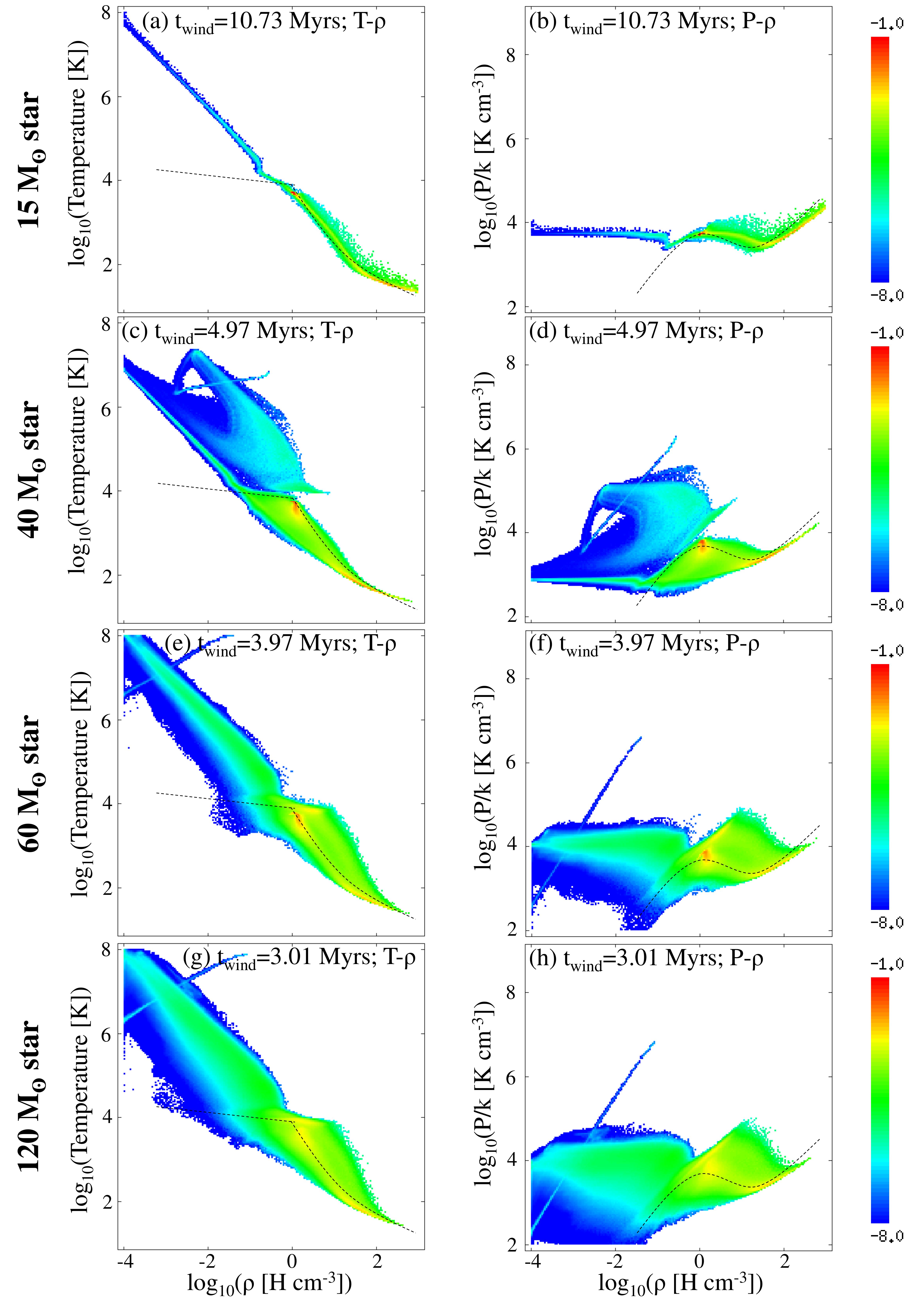}
\caption{Mass-weighted temperature-density and pressure-density distributions at the
  end of the star's life for each of the simulations. Over-plotted dashed
lines indicate the approximate thermal equilibrium between heating and cooling 
for the prescriptions used. Raw data: doi.org/10.5518/201.} 
\label{wndphases}
\end{figure*}

\begin{figure*}
\centering
\includegraphics[width=160mm]{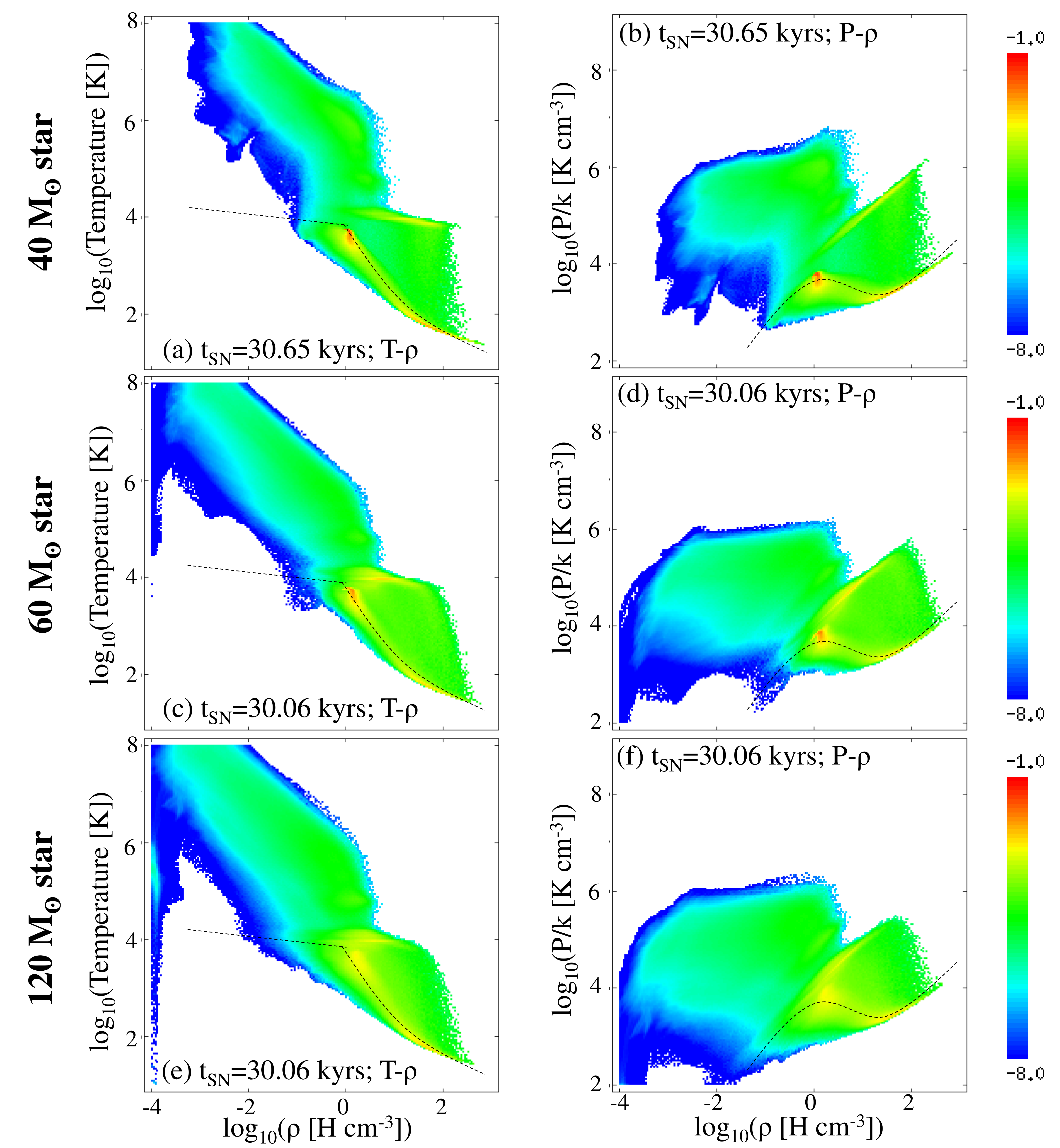}
\caption{Mass-weighted temperature-density and pressure-density distributions for the simulations
presented here during the early evolution of the supernova in each case. Over-plotted dashed
lines indicate the approximate thermal equilibrium between heating and
cooling for the prescriptions used. Raw data: doi.org/10.5518/201.} 
\label{snphases}
\end{figure*}

\begin{figure*}
\centering
\includegraphics[width=150mm]{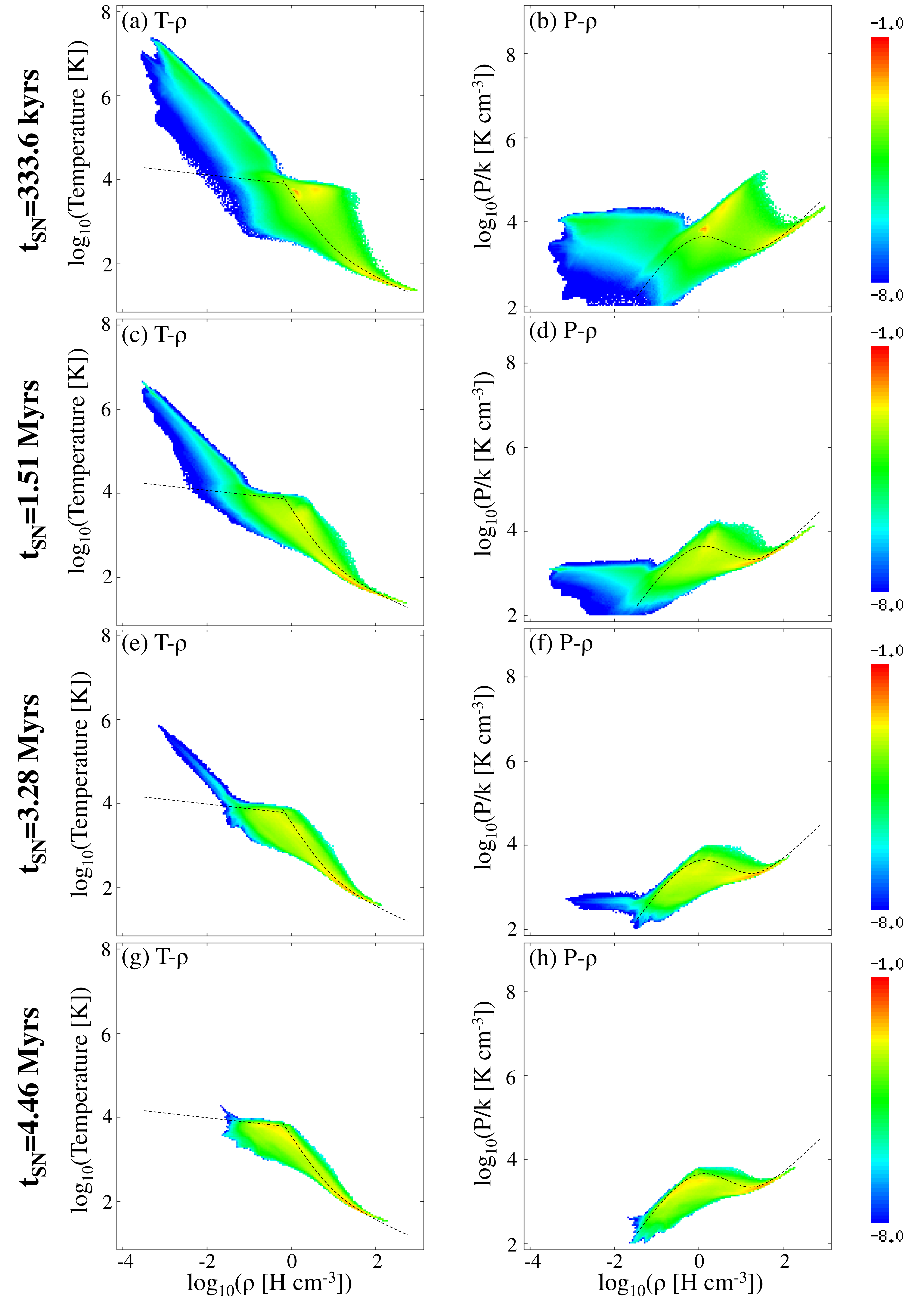}
\caption{Mass-weighted temperature-density and pressure-density distributions for the 60\,M$_{\odot}$
star case in the late stages of the supernova event. The evolution of the 40\,M$_{\odot}$
and 120\,M$_{\odot}$ star cases are very similar. Over-plotted dashed lines indicate 
the approximate thermal equilibrium between heating and cooling for
the prescriptions used. Raw data: doi.org/10.5518/201.} 
\label{snlatephases}
\end{figure*}

In Fig.~\ref{phases} we show the phase fraction and total cloud mass
for all four feedback simulations. At the injection of a star,
27\,Myrs into the cloud's evolution, the cloud is in thermal
equilibrium (i.e. on the equilibrium curve between heating and cooling
processes) with 50\% of its mass in the warm phase (5,000\,K $\leq$ T
$\leq$ 10,000\,K), $\approx 25$\% of its mass in the cold phase (T
$\leq$ 160\,K) and $\approx 25$\% of its mass in the unstable
equilibrium part of the pressure-density distribution (160\,K $\leq$ T
$\leq$ 5,000\,K). No material is in the hot thermal phase (T $\geq$
10,000\,K).

Once mechanical feedback starts the cloud material is shock heated out
of thermal equilibrium and requires a cooling time to radiate away its
excess energy. The low density shocked stellar wind gas has a
comparatively long cooling timescale, while the denser and cooler
swept-up gas has a much shorter cooling timescale. Later on, the hot
SN ejecta and the gas it sweeps up have comparatively long cooling
timescales.

Fig.~\ref{phases}(a) shows that during the wind phase of the
15\,M$_{\odot}$ star case before the cloud collapses (beyond
$t=37$\,Myrs), the fraction of cold material decreases during the
first 0.5\,Myr, with a corresponding increase in unstable material,
then slowly increases very much like the reference case with no
feedback - the molecular cloud continues to form around the stellar
wind bubble. The introduction of stellar feedback has little effect
upon the amount of warm thermally stable material, until close to the
collapse of the cloud. The presence of a larger quantity of unstable
material as compared to the reference case enables the condensation of
slightly more cold material, allowing the fraction of cold material to
increase above the reference case with feedback. One of the very few
effects of the 15\,M$_{\odot}$ star then, is to slightly increase the
amount of cold molecular material in the cloud.

In panel (b) of Fig.~\ref{phases} we show the phase evolution of the
material in the simulation of mechanical feedback from the
40\,M$_{\odot}$ star. The effect upon the cloud is greater than that
of the 15\,M$_{\odot}$ star. Whilst the warm fraction remains
reasonably constant for the first 2\,Myrs, the stellar wind rapidly
heats cold cloud material, moving it back into the unstable phase
(from where it all originally began in the diffuse initial condition
preceding the formation of the molecular cloud at t=0).  The unstable
phase fraction increases and the cold phase fraction decreases. Over
the rest of the star's life though, both the cold and unstable phase
fractions continue to grow, at the expense of the warm phase, which
drops to below 20\% by the time of the supernova, compared to 50\% in
the reference case at the same time. Clearly the star reduces the
amount of cold material formed in the cloud (to 30\%, rather than more
than 40\% in reference case), but is not able to stop the cloud
continuing to form cold material. The SN then transports hundreds and
eventually thousands of solar masses of SN, wind and then cloud
material out of the cloud. The SN rapidly heats nearly 10\% of the
cloud material into the hot thermal phase (above 10,000\,K), but the
majority of this material radiatively cools quickly and the fraction
of hot material drops back close to zero before the SNR leaves the
cloud (from about 0.5\,Myrs after the explosion). The major effect of
the supernova is to transition nearly 70\% of the material on the grid
into the thermally unstable phase, mostly from the warm phase. The
amount of cold material also decreases very rapidly post-SN, returns
briefly to nearly 30\% after 0.5\,Myrs post explosion - the point at
which the SN forward shock has progressed through the cloud - and then
decreases as the remnant evolves. These changes can be understood as
the sweeping up and ejection of material from the cloud by the
SN, whilst the cold cloud material reacts much more slowly. Some of
this behaviour is also due to the (SN) shock heating and subsequent
cooling of dense material, as found by \citet{rogers13}.  The fraction
of cold dense material remains at around 20\% throughout the post-SN
evolution, with a slight increase at late-time, matched by an increase
in warm phase material, with the unstable phase feeding into
both. Cold molecular components are still present in the simulation,
albeit at a 20\% fraction rather than the 50\% fraction seen without
feedback. They have not been destroyed by the supernova, even if the
structure of the cloud has been disrupted. It should be noted though
that this result is at the limit of applicability of these
simulations, as numerical issues are present due to the forward SN
shock having long since passed off the grid.

In the case of the 60\,M$_{\odot}$ star, we observe many similarities
to the case of the 40\,M$_{\odot}$ star. The 120\,M$_{\odot}$ star
also shows similar changes, though they are more extreme, during the
stellar wind phase, with less than half the cold phase material
present by the time of supernova, as compared to the reference case,
and less than all the other cases. An appreciable fraction of hot
thermal phase material is also present, around 5\% by the time of
supernova. After the SN, the first Myr is also similar to the
preceding cases, but after that the evolution is markedly different to
the preceding cases. There is no leveling off of the cold phase
fraction, nevermind an increase. By the end of the simulation, less
than 5\% of the material in the computational volume is in the cold
phase. The volume is completely dominated by warm phase material -
nearly 80\% of the total material present - with the remainder in the
unstable phase. This is most likely because of the combined strength
of the stellar wind and subsequent supernova which completely destroy
the parent molecular cloud. Any remaining cold molecular material has
been blown out of the computational volume.  We can therefore state
that it takes more than a 60\,M$_{\odot}$ star to completely disperse
a molecular cloud by mechanical feedback under these conditions, but
that a 120\,M$_{\odot}$ star is very efficient at this dispersal.

In Fig.~\ref{wndphases} we plot the mass-weighted temperature-density 
(in the left column) and pressure-density (in the right column) 
distributions for the four cases considered in this
work, using 200 bins in log density and 200 bins in log
temperature/pressure. We also over-plot the approximate thermal
equilibrium curve for the heating and cooling prescriptions used in
this work. In the first row, we show the distributions 10.97\,Myrs
through the wind phase for the case of the 15\,M$_{\odot}$ star,
shortly before complete cloud collapse. Most of the material remains
in thermal equilibrium, tracing the equilibrium curve. The two
distinct stable phases (warm and cold) are, as expected, in
approximate pressure equilibrium with each other. A low density,
higher temperature stellar wind bubble (as shown in the bottom row of
Fig.~\ref{wnd-15M}) is indicated by the branch stretching away from
the equilibrium curve horizontally to the left at the same P/k
$\sim10^{3.8}$\,K\,cm$^{-3}$ in the pressure-density distribution.

In the second row of Fig.~\ref{wndphases}, we show the distributions
for the case of the 40\,M$_{\odot}$ star at the end of the wind
phase. The distributions have the same components on the equilibrium
curve that are present in the case of the 15\,M$_{\odot}$ star, but
there are also other considerable regions. We can identify branches
from gas in the wind bubble and the LBV/WR shell. The lower pressure
horizontal branch at P/k$ \sim10^{3}$\,K\,cm$^{-3}$ is the wind
bubble, at notably lower pressure and temperature than the previous
case. The LBV/WR shell that formed during the last stages of stellar
evolution is responsible for the branch and arc of material out of
equilibrium at higher pressure. Specifically, the diagonal line that
spans from (log(P),\,log($\rho$)) = (6,-1) down toward the tunnel
branch at (log(P),\,log($\rho$)) = (3,-3) is the wind injection region
and region of undisturbed wind material up to the reverse shock. The
reverse shock is indicated by the jump back up in pressure to the
over-pressured (with respect to the cloud and other wind bubble
tunnel) LBV/WR shell that is itself indicated by the horizontal branch
at P/k\ $\sim10^{5}$\,K\,cm$^{-3}$ that arcs back to the equilibrium
curve, broadening as it cools and increases in density. The greater
power of the wind has also driven a broadening of the distributions,
with heated material above the equilibrium curve radiatively cooling
towards the equilibrium curve. Cooling by expansion has led to low
density, cool gas below the equilibrium curve. The 
distributions in this case are very similar to the case of the
40\,M$_{\odot}$ star in Paper II, despite the marked difference in
structures; a bubble in a clumpy cloud here, a tunnel through a
filamentary sheet there. Common to both cases and responsible for much
structure in these distributions is the LBV/WR shell
structure. Looking at the distributions, it is impossible
to discern the considerably different overall wind-cloud structure
responsible for each. It may be possible to conclude that lower
pressure wind branches are indicative of larger, less confined
structures.

The third and fourth rows of Fig.~\ref{wndphases} show the 
distributions for the cases of the 60 and 120\,M$_{\odot}$ stars and
are remarkably similar. Both contain isolated diagonal lines that are
again indicative of the wind injection regions and regions of
undisturbed WR wind material up to the reverse shock. In both cases,
the jump back up in pressure at the reverse shock is at, or just off,
the lower density axis boundary of each figure, as can in fact be seen
at low density in Figs.~\ref{wnd-60M} and \ref{wnd-120M}. The greater
power of the winds and merger of the LBV/WR material with the whole
wind bubble has created a single, broader distribution, pushing more
and more material out of equilibrium, until in the case of the
120\,M$_{\odot}$ star, most of the material is out of equilibrium.

In Fig.~\ref{snphases} we show the distributions 30\,kyrs after the SN
for the three simulations with supernovae. Towards the upper left of
each distribution is the low density, hot phase, which consists of SN
shock heated gas. It is not yet in pressure equilibrium with the warm
phase, but evolves towards this equilibrium as the simulation
progresses beyond that shown here. The distribution at low density is
reasonably wide in pressure and temperature and widens with increasing
stellar mass. The reason for this is the presence of both shock heated
and adiabatically cooled gas, as noted in a similar analysis of such
distributions influenced by feedback simulated by \cite{walch15b}. The
SN has also shock-heated an appreciable amount of the cold dense
material, spreading the distribution upwards at high density in all
three plots.

In Fig.~\ref{snlatephases} we show the mass-weighted
temperature-density and pressure-density distributions at
late times after the SN for the case of the 60\,M$_{\odot}$ star. We
can see the broad distribution is cooling (shifting downwards) in both
plots.  The material is also moving back towards thermal equilibrium,
and by 4.5\,Myrs post-SN essentially all the material is back in
thermal equilibrium. The evolution of the 40\,M$_{\odot}$ and
120\,M$_{\odot}$ star cases is very similar, albeit with lower amounts
of high density material and lower maximum densities in each case
(i.e. less material on the right of equivalent plots to those shown in
Fig.~\ref{snlatephases}).

\subsection{Total cloud mass}

\begin{figure}
\centering
\includegraphics[width=80mm]{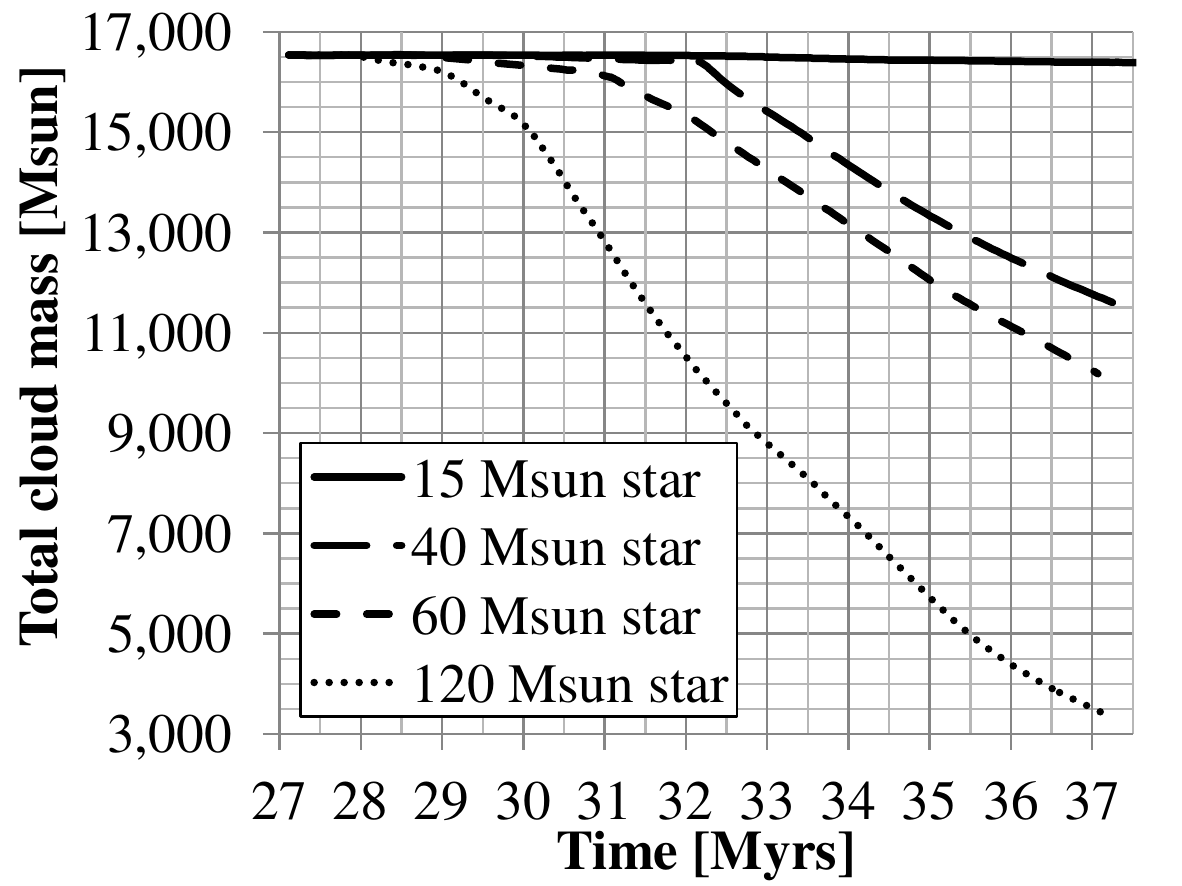}
\caption{Variation of total mass in the computational volume with time 
for the feedback simulations. The reference case without feedback is 
not shown, but almost precisely overlies the 15M$_{\odot}$ star case.
Raw data: doi.org/10.5518/201.} 
\label{cloudmass}
\end{figure}

\begin{figure}
\centering
\includegraphics[width=80mm]{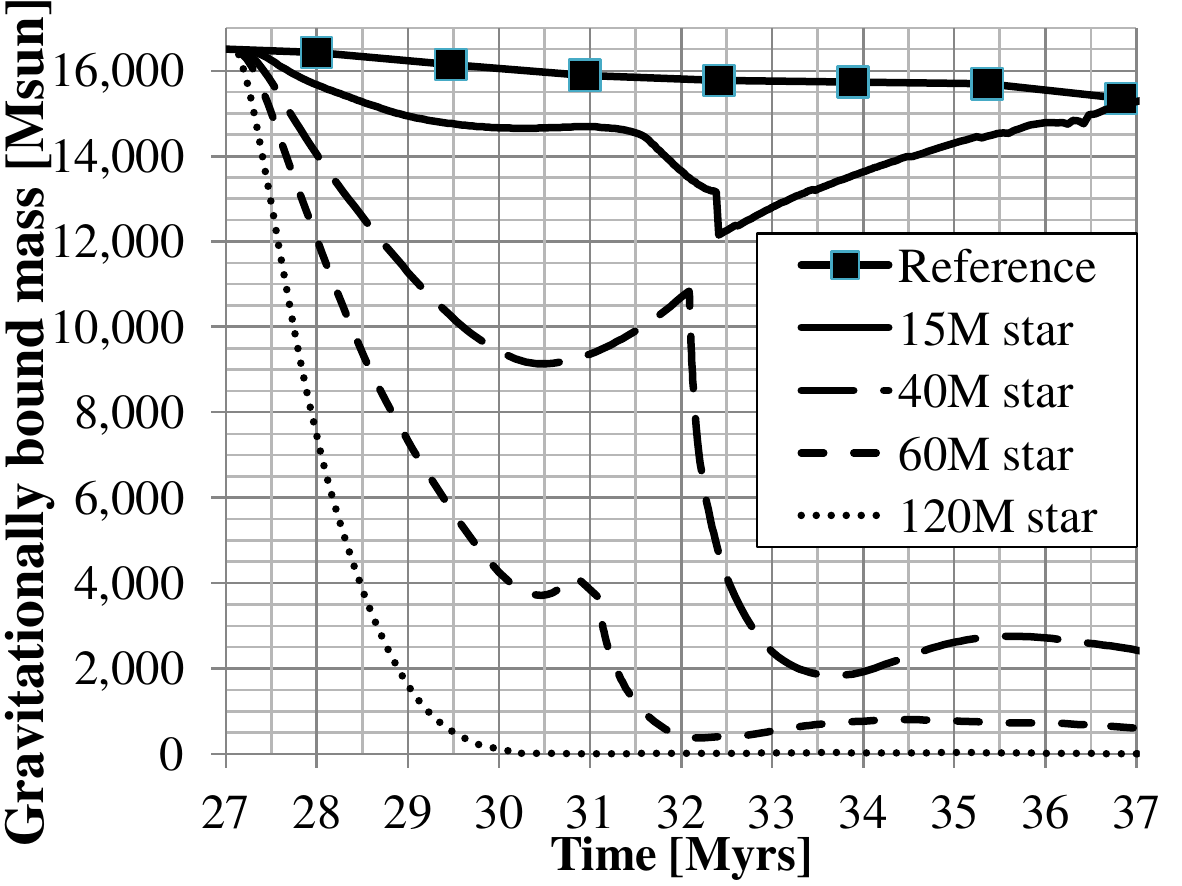}
\caption{Variation of gravitationally bound mass (i.e. kinetic energy plus
gravitational energy is less than zero) in the computational volume with time 
for the feedback simulations. Raw data: doi.org/10.5518/201.} 
\label{boundmass}
\end{figure}

\begin{figure*}
\centering
\includegraphics[width=160mm]{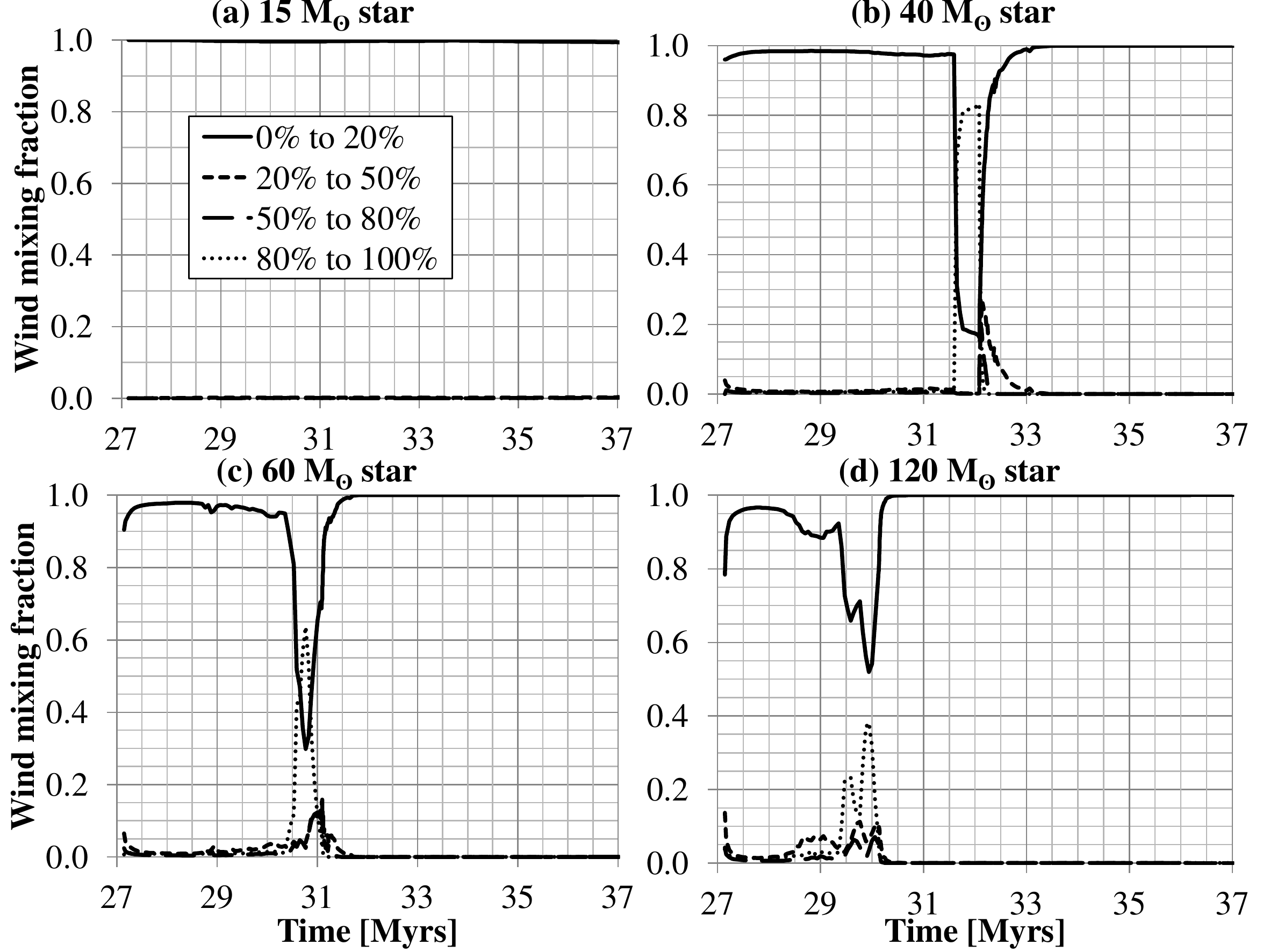}
\caption{Levels of mixing of the wind material with the cloud material 
throughout the stellar feedback simulations. A value of 100\% indicates 
no mixing of the wind material with the cloud material.
Raw data: doi.org/10.5518/201.} 
\label{windmixing}
\end{figure*}

\begin{figure*}
\centering
\includegraphics[width=160mm]{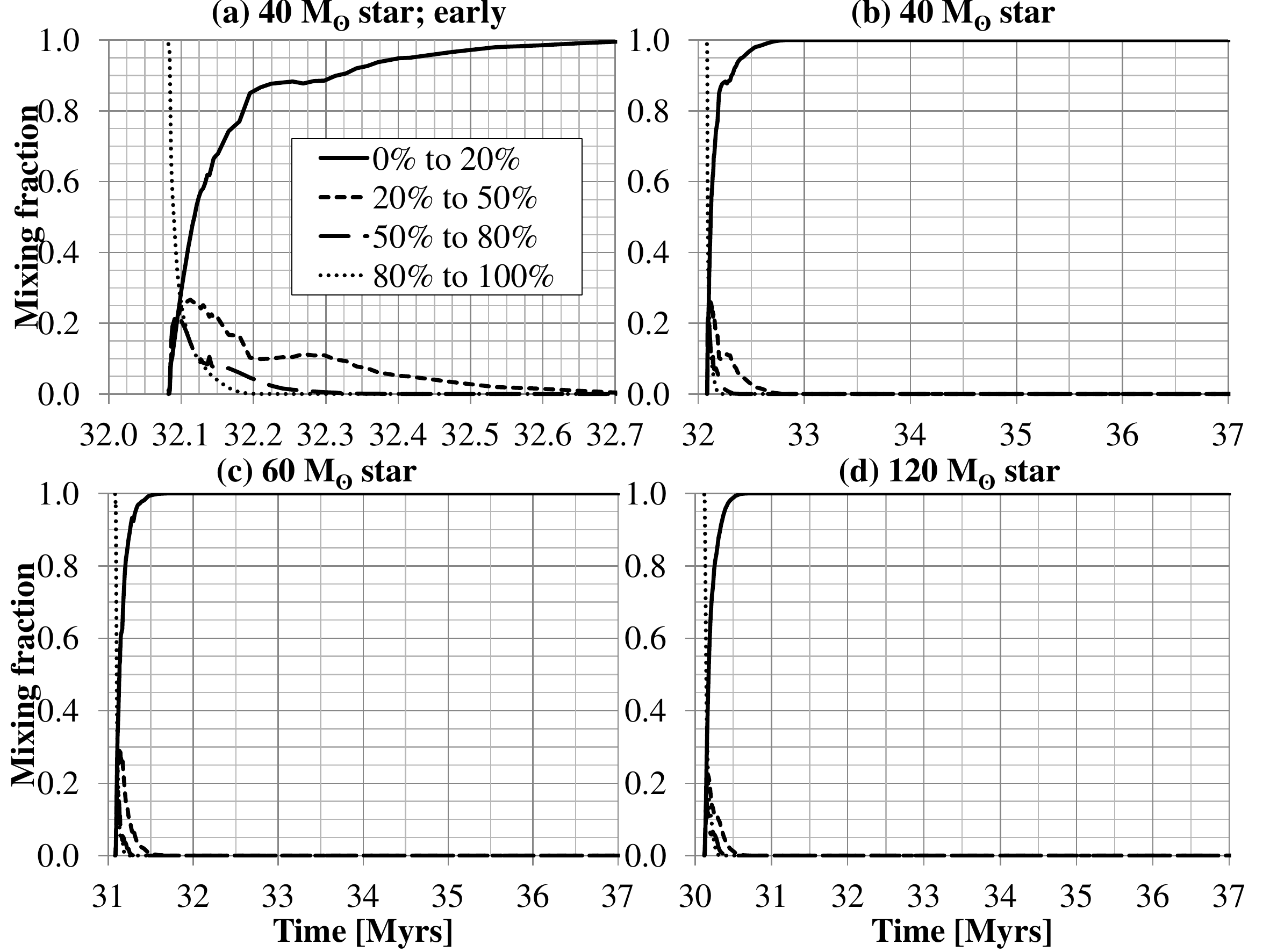}
\caption{Levels of mixing of the supernova material with the cloud/wind material 
throughout the stellar feedback simulations. A value of 100\% indicates 
no mixing of the SN material with the cloud/wind material.
Raw data: doi.org/10.5518/201.} 
\label{snmixing}
\end{figure*}

In Fig.~\ref{cloudmass} we show the variation of total mass in the computational 
volume with time for the case of each star. The reference case is not shown, as it overlaps
the 15\,M$_{\odot}$ star case, highlighting again the little influence
of the 15\,M$_{\odot}$ star. In the 40\,M$_{\odot}$ star case, the
total mass remains approximately the same until the supernova (at
32.1\,Myrs), when the SN event drives mass out of the computational 
volume. The wind phase
of the 60\,M$_{\odot}$ star drives a few hundred solar masses of
material out of the computational volume, before the supernova (at 31.1\,Myrs) expels
considerably more mass. In both the 40 and 60\,M$_{\odot}$ star cases,
the majority of mass remains on the grid post-SN until beyond
the end of the simulations. Most of it is out of equilibrium in the
thermally unstable temperature range, as revealed by the preceding
plots in this sub-section.  Even by the time of supernova (at
30.1\,Myrs), the wind from the 120\,M$_{\odot}$ star has already
driven nearly 2,000\,M$_{\odot}$ of material out of volume - more than
10\% of the cloud's original 16,500\,M$_{\odot}$.  Most of this
expulsion occurred during the stellar evolution off the Main Sequence,
during the final Myr or so of the star's life. The expulsion of mass
continues during the SN phase, until 6\,Myrs post-SN, only 20\% of the
initial mass remains in the simulation. As such, the cloud has been
effectively dispersed and a larger simulation is required to track the
final properties of this material. This would suggest that whilst the
star has destroyed the cloud and replaced it with a tenuous warm-phase
SNR surrounded by warm medium, some cold material may have survived
the violent evolution of the 120\,M$_{\odot}$ star, but is now
entirely separate from its origins in the parent molecular cloud.

In Fig.~\ref{boundmass} we show the variation of mass gravitationally 
bound to the potential of the cloud in the computational volume with time 
for the case of each star.
That is, we show the amount of mass within the gravitational potential
well of the original cloud that does not have enough kinetic energy to
escape the gravitational potential of the cloud.
Here is the only place the 15\,M$_{\odot}$ star has an appreciable
effect over the reference case, although by the end of the star's evolution,
the amount of gravitationally bound mass has returned to the same level as the
case without
feedback. In the 40\,M$_{\odot}$ star case, the main sequence evolution
of the star provides enough energy for nearly half the mass of the cloud to
overcome the gravitational energy and either leave the domain, or have
enough kinetic energy to overcome the gravitational potential, but the late stages of evolution increase
the amount of mass associated with the cloud. The effect of the supernova
is to disperse the remainder of the cloud, although 2\,Myrs after the SN, the
amount of gravitationally bound mass is increasing again, indicating that the
increasing levels of cold material noted previously are now bound to the 
cloud itself. The 60\,M$_{\odot}$ star shows a similar, but more disruptive
pattern, with the cold molecular material formed more than 2\,Myrs post-SN
now associated with the cloud. There is considerably less material in this
case. The 120\,M$_{\odot}$ star is able to
completely disperse the cloud material during the star's lifetime ($\sim$3\,Myrs), 
confirming the conclusions drawn above. No increase post-SN is noted.

\subsection{Mixing}

In Fig.~\ref{windmixing} we show the mixing of the stellar wind
material with cloud material. In the 15\,M$_{\odot}$ star case, wind
material is immediately well mixed and remains that way, since it is
all at fractions of no higher than 20\% with respect to the cloud
material. Whilst an equivalent amount of material to the star's mass
was taken out of the injection region to make the star, some cloud
material remained in the injection region and the wind material was 
introduced as a source term adding to this, hence the wind material 
mixed very quickly. There is no brief period when the
injection region is almost entirely wind material i.e. there is no
transition observed from ``no mixing" (at 100\%) down to well
mixed. Therefore, the initial few thousand years of stellar wind
may be quite effectively quenched, but this is a transitory period
and outweighed by the benefit of using a source term rather than
an imposed wind structure in those cells, which tests and previous
experience \citet[e.g.][and other works therein]{wareing07} has 
shown can impose grid-based structure upon the resulting wind.
The wind from the main sequence phase of the 40\,M$_{\odot}$
star is almost as well mixed as that of the 15\,M$_{\odot}$ star.  In
this case, the brief initial period of mixing down from almost 100\%
in the injection region is indicated by a very small initial
peak, due to the stronger stellar wind. 
However, the final 0.5\,Myrs of stellar wind injection shows the
isolated and unmixed nature of the LBV/WR shell, and also how much
material is in the shell, given the height of the peak. At
32.08\,Myrs, the SN explodes and in short time mixes the wind material
down to concentrations of below 20\% wind material. Both the 60 and
120\,M$_{\odot}$ star cases show peaks of unmixed material in the
final stages of stellar evolution, but with decreasing levels, down to
the 120\,M$_{\odot}$ star case where the majority of wind material
remains efficiently well-mixed throughout the star's lifetime, albeit
with larger fractions of other levels of mixing than in any of the
other cases. Peaks and troughs in the final stages of stellar
evolution can be attributed to the different phases of the stellar
wind and transitory shells that arise during these stages. The effect
of the supernova on the mixing of wind material is similar and is very
efficient in all cases.

In Fig.~\ref{snmixing} we show the mixing of supernova material with
cloud and wind material in the three supernova cases presented in this
work. In panel (a) we focus upon the early mixing of the supernova
ejecta, in the first few hundred thousand years following the
explosion of the 40\,M$_{\odot}$ star.  Initially all the SN material
is unmixed (80-100\%), remaining so for a few thousand years and then
is slowly mixed down into the more mixed brackets in this
figure. Within 0.1\,Myrs though, the SN material is efficiently mixed
down to 0-20\% fractions within the cloud/wind material. At later
times, the SN material continues to remain well mixed.  Mixing is more
efficient, taking less time to completely reach the 0-20\% brackets
for the more massive stars.

For these simulations, as for those in Paper II, we have not
quantitatively examined the nature of the material leaving the
grid. In the three more massive cases, the SN drives material out of
the cloud, as investigations of the phase fractions of material have
shown. Further investigations may shed more light on mass-loading and
the entrainment of material leaving the grid.

\section{Discussion}\label{discussion}

\subsection{Comparisons to previous work}

Comparing the present work with our previous efforts in the
hydrodynamic case \citep{rogers13,rogers14} is not straightforward due
to the large differences in initial conditions. However, there are
some clear similarities. In both, the later wind energy and SN energy
is transferred to the wider surroundings through a multitude of porous
channels. In Rogers \& Pittard this was as a result of the imposed
initial condition and was associated with a much higher density
filamentary cloud in a smaller volume. Here, the early phases of the
stellar wind result in similar channels and gaps, even though the
initial condition is that of a clumpy cloud with no filamentary
network. This finding, of clear gaps and channels carved out through
parental molecular clouds, would appear to be a common evolutionary
conclusion when mechanical feedback from stars is involved. Only when
a magnetic field is present, effectively focussing the outflow from a
massive star into a single channel as we found in Paper II, are the
results morphologically and quantitatively different. Nevertheless, it
is clear that winds are on the whole quite capable of breaking out of
their confining clouds. Such ‘leakage’ of the hot gas is consistent
with the much lower mass of hot cluster gas than expected for the
cluster ages and the mass-loss rates of stars \citep{townsley03} 
and by energy budget considerations \citep{rosen14}.

In common with Rogers \& Pittard, the densest molecular regions are
surprisingly resistant to ablation by the stellar/cluster wind. In
both works this is partially due to shielding by other dense regions
closer to the star/cluster. Pressure gradients within the flow appear
to rearrange the molecular cloud structure into radially-aligned
spokes or filaments, so that the head of the spoke closest to the star
shields the remainder of the structure.  Further simulations are
required to investigate whether this behaviour extends to clumpy
clouds of higher mass and density.

In common with all our previous work, SNe are able to transport large
amounts of energy directly out of the parent cloud, with existing
channels and gaps allowing weaker coupling to the remaing dense
material. Such channels are further carved out by the SN. The key
factor is the shaping effect of the pre-SN stellar winds, which make
the cloud highly porous \citep[][and the current paper]{rogers13}, or
as in Paper II where they are focussed by magnetic fields to open up
large-scale channels directing SN energy out of the cloud. In all
cases, the winds appear to be better at removing molecular material
from the cloud/cluster environment, despite typically injecting less
energy than the SN(e). The molecular material in all cases was found
to be almost completely removed from the original cloud after 6\,Myr
post-SN. These works together demonstrate the complexity of the
interaction of a stellar wind with an inhomogeneous environment. The
results are far removed from simple spherically symmetric models and
compare nicely with the simulations of \cite{geen16} who argue that
strong pre-supernova (radiative) feedback is required to allow
supernova blastwaves to propagate efficiently into the interstellar
medium.

A key common finding between this work and Paper II is that lower mass
stars (15\,M$_{\odot}$ or less) have little global effect on the
parent cloud. Here we find that the weak wind is unable to overcome
the gravitational collapse of the cloud. In Paper II, the local small
bipolar wind bubble is eventually refilled by the slow wind from the
late stages of the star's evolution, as we also note here. The effect
in both cases is as if the stellar wind had not been present at all.

\cite{harper09} considered stellar wind feedback into an inhomogeneous
environment and postulated that the non-uniform surrounding medium
causes gaps in the swept-up shell surrounding the wind-blown bubble
where some of the high-pressure gas in the bubble interior can leak
out. We have simulated a scenario very similar to this, with a stellar
wind expanding into a non-uniform, inhomogeneous clumpy cloud. For the
intermediate mass cases (stars of mass 40\,M$_{\odot}$ and
60\,M$_{\odot}$) we find exactly that - the wind shell is
non-spherical and has gaps or channels through it which allow gas to
vent or blow out of the parent cloud. In the case of the
15\,M$_{\odot}$ star, this effect is far less pronounced - the bubble
which forms remains close to the star (within 10 or so parsecs) and no
gaps or channels are seen around or outside the bubble. At the other
extreme, the wind-bubble around the 120\,M$_{\odot}$ star destroys
half the cloud by the end of the wind phase, and channels are present
through the remaining cloud. In this case, dramatic structural changes
occur within the cloud near the end of the star's life due to the
increasingly powerful wind blowing at this time.

The effect of stellar winds on molecular clouds has also been
considered by \cite{dale15} and \cite{offner15}. These simulations
used either momentum driven or isothermal winds and so give a lower
limit to their impact. With this is mind, \cite{offner15} concluded
that stellar mass-loss rates for individual stars must be greater than
10$^{-7}$\,M$_\odot$\,yr$^{-1}$ in order to reproduce shell
properties. As we noted in Paper II, the mass-loss rates for the
15\,M$_{\odot}$ star are less than
$3 \times 10^{-7}$\,M$_\odot$\,yr$^{-1}$ this limit for the entire MS
evolution of the star (see figure 2 of Paper II). When the mass-loss
rate exceeds this limit the star is in its RSG phase, so the wind is
slow and has little power.  On the other hand, the higher mass stars
always have mass-loss rates above this limit (as shown in
Figs.~\ref{60evolution} and \ref{120evolution} of the current paper
and figure 3 of Paper II). Therefore our findings are not in
disagreement with \cite{offner15}, although we have not tested stars
with masses between 15\,M$_{\odot}$ and 40\,M$_{\odot}$. The wind from
the 15\,M$_{\odot}$ star has little effect on the parent cloud,
restricted to a 10\,pc radius around the star. In contrast, the winds
from higher mass stars strongly affect the parent molecular cloud.

Numerous recent works have explored the effects of supernovae on the
multi-phase ISM and parent molecular clouds
\citep[e.g.][]{gatto15,walch15,walch15b,giri16,kortgenphd,kortgen16}.
However, there is still much disagreement on the ability of supernovae
to drive outflows.  For instance, \cite{giri16} found that strong
outflows were only generated when SNe were randomly positioned and had
the opportunity to inject energy into relatively low density
environments. SNe placed at density peaks instead radiate away too
much energy to drive any noticeable outflow. However, they did not
account for the effect of the SN-preceding stellar wind. In contrast,
\cite{simpson16}, who accounted for the production of cosmic rays in
supernova events, showed that outflows can be driven from SNe placed
at density peaks. They found their outflows have similar mass loading
as obtained from random placement of SNe with no cosmic rays.
\cite{kortgen16} find that single SNe disperse $\sim$10\,pc-sized
regions but do not disrupt entire clouds, which instead requires
clustered, short-interval SNe to form large hot bubbles. This is in
contrast to our findings, but in mitigation our simulations focus on
feedback in a cloud of lower mass than the \cite{walch15} and
\cite{kortgen16} studies, and also include the effect of the stellar
wind, so it is not unexpected that our conclusions are somewhat
different. Compared to these other works, our simulations clearly show
that when stars are placed at high density locations, their stellar
winds can create gaps and carve channels out of the clumps and the
parent cloud, allowing wind material and SN material to escape easily 
from the cloud, mass-loaded with material stripped from clumps in the
parent cloud (\cite{fierlinger16} also find that pre-SN feedback
enhances the impact of SNe). Thus the prior influence of winds should
be included in simulations involving SN feedback.

\subsection{Comparison to observations}

\cite{lopez11} support the scenario presented by \cite{harper09},
concluding that leakage through gaps in the swept-up wind-blown bubble
may be occurring within 30 Doradus. They claim that the lower X-ray
gas pressure relative to the direct radiation pressure suggests that
the hot gas is only partially confined and is hence leaking out of
``pores'' in the H{\sc ii} shell. They also conclude that the
significant radiation pressure near the central star cluster indicates
that direct stellar radiation pressure dominated the interior dynamics
at earlier times, but this claim has proved far more controversial,
and other works are in favour of the thermal pressure of hot X-ray
emitting plasma shaping the large-scale structure and dynamics in 30
Doradus. For example, \cite{pellegrini11} find that the dynamics and
large-scale structure of 30 Doradus are set by a confined system of
X-ray bubbles in rough pressure equilibrium with each other and with
the confining molecular gas. In both the 60 and 120\,M$_{\odot}$ cases
presented herein, we find a structure of wind and cloud material in
rough pressure equilibrium, as shown in Fig.~\ref{wndphases}. The
existence of low-density channels carrying hot wind material is also
clear from Figs.~\ref{wnd-40M} and \ref{wnd-120M}. The more powerful
the wind, the larger the cavity, the quicker the channels form and the
wider they are, implying that a cluster of stellar winds will rapidly
form multiple wide channels through a low-mass parent molecular cloud
such as studied here. However, to determine whether hot gas thermal
pressure or direct radiation pressure is dominant requires simulations
with both mechanical and radiative feedback included.

\cite{lopez14} further assessed the role of stellar feedback at
intermediate scales of 10-100\,pc. They studied a sample of 32 H{\sc
  ii} regions (with ages from 3-10\,Myr) in the Large and Small
Magellanic Clouds and found that warm ionized gas dominates the
dynamics whilst the hot gas pressures are comparatively weak. They
emphasize that younger, smaller H{\sc ii} regions should be studied to
elucidate the roles of hot gas and direct pressure in the early
evolution of these regions. They also conclude that the hot gas is
only partially confined in all of their sources, and suggest that hot
gas leakage is common. Our simulations are by design influenced only
by hot gas pressure, but nevertheless create the kind of structures
observed by \cite{lopez14} and others.

It is also interesting to note the existence of structures within
H{\sc ii} regions. For example, in RCW 120 arcs of dust emission are
clearly seen within the H{\sc ii} region bubble in mid-IR {\it
  Spitzer} data. \citep{mackey16} interpret this as the outer edge of
the wind-blown-bubble. An alternative viewpoint is that dust arcs are
waves induced by photo-evaporation flows inside H{\sc ii} bubbles
\citep{ochsendorf14}.  Our simulations, although not able to define
the extent of the H{\sc ii} region, do show complex structure within
the wind bubble itself, for example bow shocks and LBV/WR shells,
which could perhaps also lead to such observed structures. Such arcs
could also be related to the aligned radial structures formed in the
simulation of the 60\,M$_{\odot}$ star. Photoionisation of the complex
structures formed by the 60 and 120\,M$_{\odot}$ stars is bound to
show one-sided H{\sc ii} regions too, as the stellar wind has blown
out the molecular cloud in unequal measure around the star, as shown
in Figure \ref{wnd-proj}.

Some H{\sc ii} regions (e.g. RCW 79, RCW 82, RCW 120) show a central
hole in the 24\,$\mu$m emission \citep[e.g.][]{martins10}, which may
also be evidence of stellar winds. However, these holes could also have been
produced by radiation pressure. Simulations by
\citet{freyer03,freyer06} show that stellar winds can have important
dynamical effects even when the ratio of the injected wind to ionizing
photon energy is as low as 0.01, but it is not known if this remains
true at still lower ratios.

Two ``smoking guns'' which reveal strong evidence for the potential
impact of stellar winds are bowshocks around stars on the periphery of
stellar clusters \citep[e.g.][]{winston12}, and diffuse X-ray emission within
and around young pre-SN clusters
\citep[e.g.][]{gudel08,townsley14}. However, it is clear that further
observational and theoretical work is needed for a better
understanding of the dynamics of H{\sc ii} regions and
wind-blown-bubbles/cluster outflows.

\section{Summary and conclusions}\label{conclusions}

In this work we have explored the effects of mechanical stellar wind
and supernova (SN) feedback on realistic molecular clouds without
magnetic fields. Our initial condition has been based on the work of
\cite{wareing16} in which a diffuse atomic cloud was allowed to form
structure through the action of the thermal instability, under the
influence of gravity, but without injected turbulence. The resulting
structure is best described as a clumpy near-spherical cloud of
approximately 100\,pc-diameter surrounded by a diffuse atomic cloud,
when we introduce mechanical stellar feedback. The clumps are roughly
equally distributed over the inner region of the cloud (r$\leq$50\,pc)
when the densities in some clumps reach 100 cm$^{-3}$. A single
massive star was then introduced at the highest density location
closest to the centre of the cloud. We considered four cases:
formation of a 15\,M$_{\odot}$ star, a 40\,M$_{\odot}$ star, a
60\,M$_{\odot}$ star and a 120\,M$_{\odot}$ star. Their stellar winds
(based on realistic Geneva non-rotating stellar evolution models)
subsequently affect the parent cloud, and at the end of life of the
three highest mass stars a SN explosion is modelled. We do not model a
SN explosion in the case of the 15\,M$_{\odot}$ star as the cloud
collapses under gravity before the end of the star's life and the
simulation loses meaning at the resolution available.

In the 15\,M$_{\odot}$ star case the stellar wind has very little
effect on the molecular cloud, and forms only a small cavity around
the star which reduces in size as the cloud collapses and the star
enters the final stages of stellar evolution. The stellar wind is
unable to support the wind-blown bubble against the surrounding
gravitational collapse of the cloud.

In the more massive star cases, the stellar wind has a significant
effect on the molecular cloud, carving channels and gaps through the
cloud which are reminiscent of the interaction seen in our previous
work \citep{rogers13,rogers14}.  Each star's wind halts the
gravitational collapse of the cloud.  The density, temperature,
pressure and velocity of material in the cloud environment all span
many orders of magnitude. The hottest gas typically occurs at the
reverse shock of the stellar wind, and cools as it expands away from
the cloud and mixes in with denser surrounding material. A multitude
of weaker shocks form around the remaining dense clumps within the
cloud and material is ablated from these into the stellar wind flow,
as they also rearrange and form radially aligned spokes in the cloud.
This change of structure allows self-shielding to occur and for cold
molecular cloud material to survive for the lifetime of the star, both
in these shielded spokes and entrained into the wind escaping the
cloud along the channels.

In the most massive (120\,M$_{\odot}$) star case, nearly half the
molecular cloud is dispersed and destroyed during the final stages of
stellar evolution, before the SN occurs. The channels through the
wind bubble allow the forward shock of the SN to leave the wind
structure very quickly through these 'leaky gaps'. In the cases of both 
the 60\,M$_\odot$ and 120\,M$_\odot$ stars, along these wind-carved
channels the forward shock leaves the computational volume after only 
15,000 years, at an average speed of over 4,500 km\,s$^{-1}$, efficiently
transporting kinetic and thermal energy into surroundings beyond the
parent molecular cloud. The SN subsequently continues
this dispersal and destruction, overrunning the structures formed
during the wind phase. In all three cases, by 2.5\,Myrs after the SN,
material with the characteristics of the warm neutral medium surrounds
a low-density inner cavity (the central region of the SNR) at the
original location of the isolated star.

It can be concluded from this work that stellar winds from the lower
mass end of the range of stars undergoing core-collapse have little
effect on their parent cloud. In such cases, the cloud is likely to
evolve subject to gravity, radiation pressure, and external
influences, until the star explodes. In contrast, higher mass stars
are able to disperse and destroy the cold molecular material of the
parent cloud even before SNe occur. Massive stars of intermediate mass
carve channels and gaps through their parent clouds, allowing SN
material to remain fairly hot and energetic as it escapes \citep[see
also][]{rogers13,rogers14,wareing17}.

In this work, as in \citet{wareing17}, we have taken a single initial
condition, realistically formed from the action of the thermal
instability, but none-the-less in isolation and at the lower end of
molecular cloud masses. Molecular cloud masses in the Milky Way reach
10$^{5-6}$\,M$_{\odot}$, considerably more than the
16,500\,M$_{\odot}$ of material in the cloud investigated here. The
nature and distribution of this material is key to how the stellar
winds and SN affect the molecular cloud. In future work we will
examine feedback into higher mass clouds and also account for
radiative feedback effects.

\section*{Acknowledgments}

This work was supported by the Science \& Technology Facilities
Council [Research Grant ST/L000628/1]. The calculations for this paper
were performed on the DiRAC Facility jointly funded by STFC, the Large
Facilities Capital Fund of BIS and the University of Leeds and on
other HPC facilities at the University of Leeds. These facilities are
hosted and enabled through the ARC HPC resources and support team at
the University of Leeds (A. Real, M.Dixon, M. Wallis, M. Callaghan \&
J. Leng), to whom we extend our grateful thanks.  We acknowledge
useful discussions with T. W. Hartquist and S. Van Loo and extend
further thanks to S. Van Loo for the provision of analysis routines to
produce PDFs.  We also express thanks to the Reviewer, Prof. Vincent
Icke, and Editors for their comments on the first draft of this manuscript. 
Raw data for the all figures in this paper is available
from the University of Leeds Repository at http://doi.org/10.5518/201.  
VisIt is supported by the Department of
Energy with funding from the Advanced Simulation and Computing Program
and the Scientific Discovery through Advanced Computing Program.

\label{lastpage}

\end{document}